\documentclass[11pt,a4paper]{article}

\usepackage{a4wide}
\usepackage{epsfig}
\begin{document}
\setcounter{page}{0}
\thispagestyle{empty}
\noindent
{\large UW-PT-98-06}\hfill{\large hep-th/9805124}
\vspace{0.4cm}

\begin{center}
\huge{\bf
Aspects of Quasi-Phasestructure\\
of the Schwinger Model on a Cylinder\\
with Broken Chiral Symmetry}
\end{center}
\vspace{0.6cm}
\begin{center}
{\large\bf Stephan D\"urr}\\
\vspace{0.2cm}
{\sl University of Washington\\
Particle Theory Group, Box 351560\\
Seattle, WA 98195 (U.S.A.)}\\
\vspace{0.1cm}
{\large\tt durr@phys.washington.edu}
\end{center}
\vfill
\begin{abstract}
\noindent
We consider the $N_f$-flavour Schwinger Model on a thermal cylinder of 
circumference $\beta\!=\!1/T$ and of finite spatial length $L$. On the
boundaries $x^1\!=\!0$ and $x^1\!=\!L$ the fields are subject to an element of
a one-dimensional class of bag-inspired boundary conditions which depend on a
real parameter $\theta$ and break the axial flavour symmetry. For the cases
$N_f\!=\!1$ and $N_f\!=\!2$ all integrals can be performed analytically. While
general theorems do not allow for a nonzero critical temperature, the model is
found to exhibit a quasi-phase-structure: For finite $L$ the condensate -- seen
as a function of $\log(T)$ -- stays almost constant up to a certain temperature
(which depends on $L$), where it shows a sharp crossover to a value which is
exponentially close to zero. In the limit $L\to\infty$ the known behaviour for
the one-flavour Schwinger model is reproduced. In case of two flavours direct
pictorial evidence is given that the theory undergoes a phase-transition at
$T_c\!=\!0$. The latter is confirmed --~as predicted by Smilga and
Verbaarschot~-- to be of second order but for the critical exponent $\delta$
the numerical value is found to be 2 which is at variance with their
bosonization-rule based prediction $\delta\!=\!3$.
\end{abstract}
\clearpage


\newcommand{\pas}{\,\partial\!\!\!/}
\newcommand{\psl}{\,\partial\!\!\!/}
\newcommand{\Dsl}{D\!\!\!\!/}
\newcommand{\lrar}{\longrightarrow}
\newcommand{\pa}{\partial}
\newcommand{\psb}{\overline{\psi}}
\newcommand{\psd}{\psi^{\dagger}}
\newcommand{\etd}{\eta^{\dagger}}
\newcommand{\chd}{\chi^{\dagger}}
\newcommand{\tr}{\,{\rm tr}\,}
\newcommand{\Tr}{\,{\rm Tr}\,}
\newcommand{\til}{\tilde}
\renewcommand{\dag}{^\dagger}
\newcommand{\pri}{^\prime}
\newcommand{\pr}{\prime}
\newcommand{\ha}{{1\over 2}}
\newcommand{\hb}{\hbar}
\renewcommand{\>}{\rangle}
\newcommand{\ran}{\rangle}
\newcommand{\<}{\langle}
\newcommand{\lan}{\langle}
\newcommand{\gaf}{\gamma_5}
\newcommand{\lap}{\triangle}
\newcommand{\paw}{\par} 
\newcommand{\pan}{\newline}
\newcommand{\uad}{\ }
\newcommand{\al}{\alpha}
\newcommand{\be}{\beta}
\newcommand{\ga}{\gamma}
\newcommand{\de}{\delta}
\newcommand{\ep}{\epsilon}
\newcommand{\ve}{\varepsilon}
\newcommand{\ze}{\zeta}
\newcommand{\et}{\eta}
\renewcommand{\th}{\theta}
\newcommand{\vt}{\vartheta}
\newcommand{\io}{\iota}
\newcommand{\sg}{\sgppa}
\newcommand{\la}{\lambda}
\newcommand{\rh}{\rho}
\newcommand{\vr}{\varrho}
\renewcommand{\sg}{\sigma}
\newcommand{\ta}{\tau}
\newcommand{\ph}{\phi}
\newcommand{\vp}{\varphi}
\newcommand{\ch}{\chi}
\newcommand{\ps}{\psi}
\newcommand{\om}{\omega}
\newcommand{\ov}{\over} 
\newcommand{\cd}{{\cal D}}
\newcommand{\rch}{{\rm ch}}
\newcommand{\rsh}{{\rm sh}}
\newcommand{\msh}{\mbox{sh}}
\newcommand{\mch}{\mbox{ch}}
\newcommand{\mssh}{\mbox{\small sh}}
\newcommand{\msch}{\mbox{\small ch}}
\newcommand{\beq}{\begin{equation}}
\newcommand{\eeq}{\end{equation}}
\newcommand{\bdm}{\begin{displaymath}}
\newcommand{\edm}{\end{displaymath}}
\newcommand{\bea}{\begin{eqnarray}}
\newcommand{\eea}{\end{eqnarray}}
\newcommand{\bes}{\begin{eqnarray*}}
\newcommand{\ees}{\end{eqnarray*}}

\normalsize


\section{Introduction}


For realistic gauge field theories like QCD(4) it is in general an unsolved
problem to determine their phase-structure (e.g. as a function of the fermion
masses $m_{1}\ldots m_{N_f}$) analytically.
For this reason, one may either determine their phase-structure approximatively
or try to attack the question in some simpler models analytically
\cite{ShBook,SmBook}.

For many questions arizing in QCD, the Schwinger model \cite{SMorg} (QED in two
dimensions with one or $N_f$ massless fermions) has proven to be an interesting
testing ground:
It has shed some light on such longstanding problems of QCD as the
$U(1)_A$-problem \cite{FrAdHeHoIs} and it has been used to test the validity
of the Instanton Liquid Picture \cite{InstantonLiquid,SmSm}.

However, for investigating symmetry-breakdown any two-dimensional model field
theory doesn't seem to be of interest: There is no spontaneous breaking of a
continuous global symmetry with associated Goldstone bosons in two dimensions
\cite{Coleman} and a symmetry which is anomalously broken can't get restored
at finite temperature \cite{DoJa}.

It is the aim of the present paper to show that --~in spite of the truth of
this conventional wisdom~-- the Schwinger model exhibits an interesting
quasi-phase-structure:
The chiral condensate $\<\psd\ha(1\pm\gaf)\psi\>$ (which is used to probe the
chiral symmetry) shows --~as a function of the log of the temperature~-- a sharp
crossover-behaviour:
For any finite box-length $L$ there is a well defined low-temperature regime
where the condensate stays almost constant (the value depends on $L$) and there
is a ``critical temperature'' where the condensate decays (through a fairly well
localized symmetry-quasi-restoration process) to a value which is exponentially
close (but not equal) to zero.

\begin{figure}
\begin{tabular}{lr}
\epsfig{file=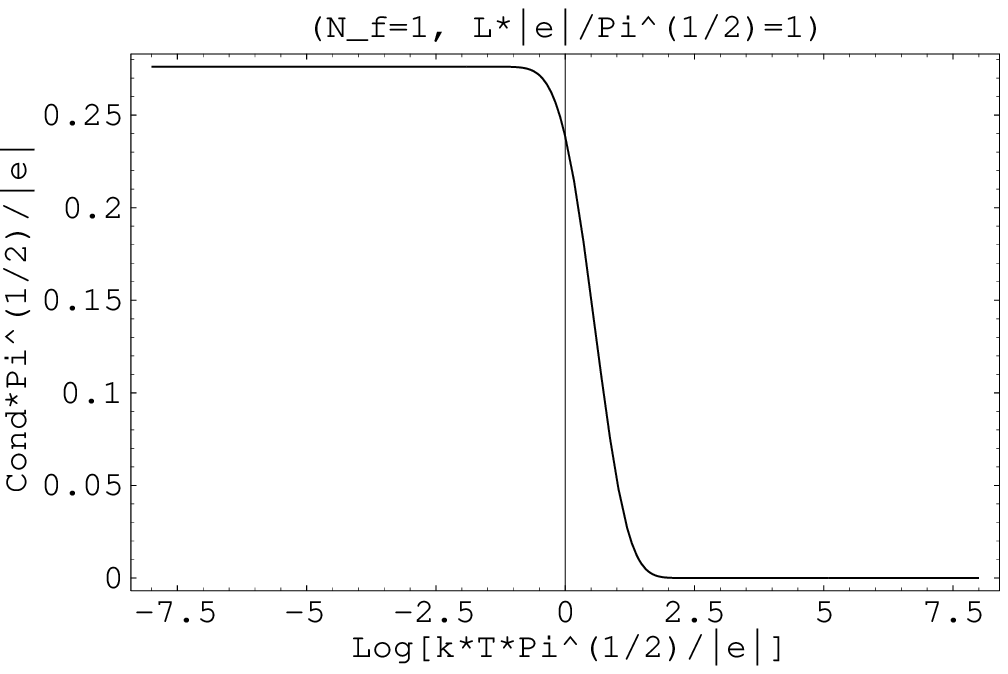,width=7cm}&
\epsfig{file=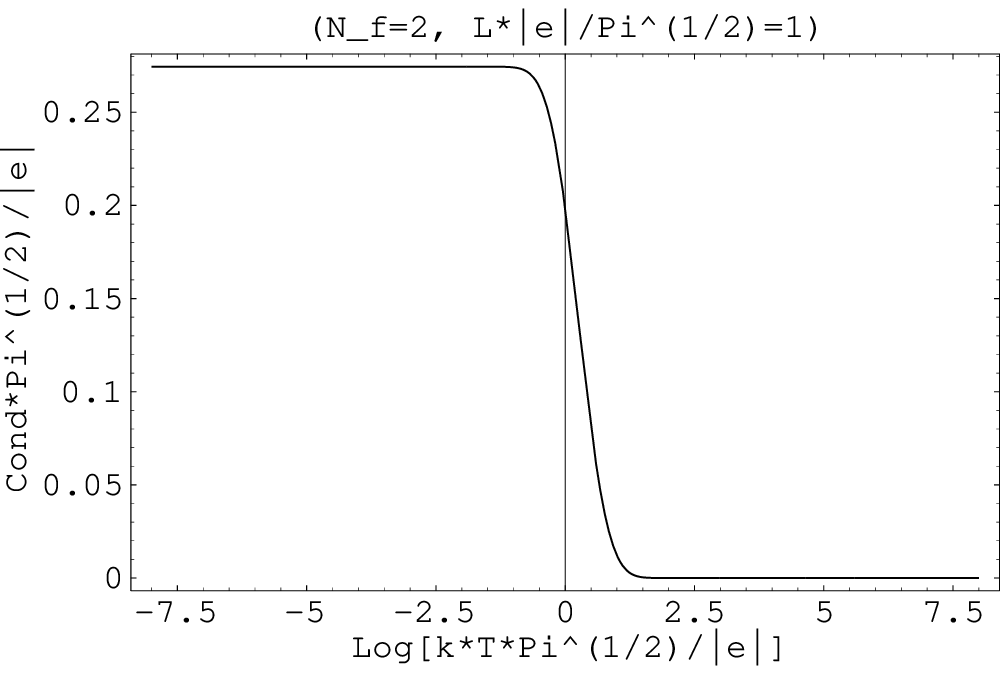,width=7cm}
\end{tabular}
\caption{The (dimensionless) condensate $\vert\<\psi\dag P_\pm\psi\>\vert/
\mu_1$ as a function of $\log(kT/\mu_1)$ at fixed box-length $L=1/\mu_1$
(where $\mu_1\!=\!|e|/\pi^{1/2}$ is the induced single-flavour Schwingermass
(\ref{schwingermass})) for $N_{\!f}=1$ and $N_{\!f}=2$.}
\label{fig1}
\end{figure}

Moreover, as the length $L$ is sent to infinity, this behaviour is shown to
provide direct evidence that the two-flavour Schwinger model exhibits a
phase-transition at $T_c\!=\!0$ and that the latter is of second order -- which
is an assumption-free rederivation of a recent claim by Smilga and Verbaarschot
\cite{SmVe}.

The Schwinger model has had a great impact on the development of field-theoretic
ideas and techniques:
The original quantization on the plane \cite{SMcon} suffered from the deficit
that a direct calculation of the condensates $\<\psd\psi\>$ and
$\<\psd\gaf\psi\>$ gave vanishing results whereas an indirect determination
via the clustering theorem led to the standard nonzero value \cite{GaSe}.
One decade ago the Schwinger model has been quantized on compact manifolds
without boundaries, the first ones being the (euclidean) sphere $S^2$
\cite{Jaye} and the (euclidean) torus $T^2$ \cite{SaWi}.
In either case direct calculations for the chiral condensates $\<\psd\psi\>$
and $\<\psd\gaf\psi\>$ were found to yield nonvanishing results for a finite
volume of the manifold.

Here we shall consider the model on a finite-temperature-cylinder with
(anti-)periodic boundary conditions in the euclidean timelike direction,
but with noncyclic boundary conditions in the spatial direction: On the two
spatial ends (at $x^1\!=\!0$ and $x^1\!=\!L$) some chirality-breaking (XB-)
boundary-conditions are imposed.

The motivation to use these XB-boundary-conditions stems again from QCD.
There one would like to give a proof (without involving any assumption) that in
the chiral limit the axial flavour symmetry $SU(N_{\!f})_A$ is spontaneously
broken.
The natural way to study a symmetry which is expected to be broken spontaneously
is to break it explicitly and to try to determine how the system behaves in the
limit when the external trigger is softly removed.
For both QCD and the $N_{\!f}$-flavour Schwinger model this means that one has
to break the axial $N_{\!f}$-flavour symmetry and then try to determine how
some observables which are sensitive to its breaking do behave in the limit
where the trigger term is removed.
There is a long series of efforts in the literature which try to achieve this
goal by studying either QCD or the $N_{\!f}$-flavour Schwinger model with a
small fermion mass.
In this approach the task is to determine how the chiral correlators behave in
the limit where the fermion masses tend to be tiny as compared to the intrinsic
mass of the theory.
The problem with this particular method of breaking the axial flavour symmetry
is that the value of the chiral condensate is related to the mean level density
of the eigenvalues of the Dirac operator in the infrared \cite{BanksCasher},
but the spectral density of the massive Dirac operator in a given gauge-field
background is not (yet) known in general.

Together with A.Wipf, we have previously explored the alternative of breaking
the chiral symmetry through introducing boundary conditions for the fermions
rather than giving them a mass \cite{WiDu}. 
There we dealt with euclidean $U(N_c)$ and $SU(N_c)$ gauge-theories with
$N_{\!f}$ massless flavours quantized inside an even($d\!=\!2n$)-dimensional
ball $B^{2n}_R$ with boundary $S^{2n-1}_R$ on which the XB-boundary-conditions
considered in \cite{HrBa} have been imposed.
These boundary conditions relate the different spin-components of each flavour
on the boundary and are neutral with respect to vector-flavour-transformations.
As a consequence the (gauge-invariant) fermionic determinant is the same for
all flavours.
The most important result of this work was the observation that the
XB-boundary-conditions proved to be equally suited to trigger a chiral
condensate as small fermion masses are.
From a technical point of view, the XB-boundary-conditions turned out to be
very convenient: In abelian theories, the (massless) fermionic determinant can
be calculated in an arbitrary gauge-field background (the fields subject to
the XB-boundary-conditions) and thus we derived in two dimensions the
analytical expressions for the chiral correlators (exploiting the rotational
symmetry of $B^2$).

In a second work \cite{DuWi}, we examined whether the approach of breaking the
$SU(N_{\!f})_A$-symmetry by boundary-conditions can be extended to
gauge-systems at finite temperature.
Choosing the Schwinger model as a testbed the answer was in the affirmative:
First of all we found that most of the nice features associated to the
quantization on contractible manifolds (topologically equivalent to $B^{2n}$)
persist in the case of the finite temperature cylinder with
XB-boundary-conditions at the spatial ends:
The configuration space is still {\em topologically trivial} (i.e. without
disconnected instanton sectors) and in particular there are {\em no fermionic
zero-modes} (which usually tend to complicate the quantization considerably
\cite{SaWi}).
The technical difficulty we had to deal with is the fact that on
non-contractible manifolds (e.g. a cylinder) the standard
decomposition-technique for the Dirac operator, on which the quantization
adopted in \cite{WiDu} heavily relied, could no longer be used. 
For this reason \cite{DuWi} turned out to be a rather technical paper, mostly
devoted to show how this difficulty can be overcome.

The present paper concentrates on the physics which can be addressed from
this setup. To be definite: The Schwinger model
\beq
\begin{array}{c}
S[A,\psd,\ps]=S_B[A]+S_F[A,\psd,\ps]
\\
\\
S_B={1\ov 4}\int\limits_M  F_{\mu\nu} F_{\mu\nu}\quad,\quad
S_F=\sum\limits_{n=1}^{N_{\!f}} \int\limits_{M}\psd_n i\Dsl\;\ps_n
\end{array}
\label{mfsm.1}
\eeq
in $d=2$ dimensions is studied on the manifold
\beq
M=[0,\be]\times[0,L]\quad\ni\quad(x^0,x^1)
\label{pro.1}
\eeq
with volume $V=\be L$.
In euclidean time direction, the fields $A$ and $\psi$ are periodic and
antiperiodic respectively with period $\beta$. Hence $x^0=0$ and $x^0=\beta$
are identified (up to an eventual minus sign) and the manifold is a cylinder.
At the spatial ends of the cylinder (at $x^1=0$ and $x^1=L$) specific
XB-boundary-conditions (to be discussed below) are imposed.
Through explicit numerical evaluation of the analytical formulae we demonstrate
that the model exhibits a quasi-phase-structure as described above.
In particular we find that the two-flavour Schwinger model (limit
$L\rightarrow\infty$) exhibits a (true) phase-transition at $T_c\!=\!0$ and
that the latter is of second order -- which is a rederivation of a result
which was obtained by Smilga and Verbaarschot \cite{SmVe} using completely
different technical tools.

For notational simplicity we use the abbreviations
\beq
\tau={\be\ov 2L}\qquad,\qquad \xi={x^1\ov L}
\label{abbrev}
\eeq
as well as the dimensionless inverse temperature and box-length 
\beq
\sg\equiv{|e|\be\ov\sqrt{\pi}}\qquad,\qquad\la\equiv{|e|L\ov\sqrt{\pi}}
\label{natuni}
\eeq
which are built from $\be, L$ by rescaling them with the Compton-wavelength
associated to the single-flavour Schwinger-mass $\mu_1$ where
\beq
\mu\;:\;\equiv\;\sqrt{N_{\!f}e^2\ov\pi}
\label{schwingermass}
\eeq
is the generalization of $\mu_1$ to the case of $N_{\!f}$-flavours.

This paper is organized as follows:
In section 2 we briefly review the quantization of the model subject to the
XB-boundary-conditions.
In section 3 the analytical results for the chiral condensate
$\<\psd\ha(1\pm\gaf)\ps\>(x)$ are re-expressed using theta-functions.
In section 4 the remaining (c-number-)integrals within the analytic formulae
for arbitrary $N_{\!f}$ are fully performed for the cases $N_{\!f}\!=\!1$ and
$N_{\!f}\!=\!2$.
This allows us to show explicit numerical evaluations of our analytical
findings for these two cases revealing the two quasi-phases as described
above and resulting in an explicit plot of the ``quasi-phase-structure''
figure \ref{fig7}.
In the concluding section 5 we try to illustrate our results and to relate
them to the work by Smilga and Verbaarschot \cite{SmVe} and others.


\section{Quantization with Thermal and XB-Boundary-Conditions}


Here we shall give a short review of the conceptually relevant aspects of the
quantization. For technical aspects the reader is referred to \cite{DuWi}.


\subsection{Chirality Breaking Boundary Conditions}

The chirality-breaking (XB-)boundary-conditions as discussed in \cite{HrBa}
can be motivated by the request that the Dirac operator $i\Dsl$ is symmetric
under the scalar product $(\ch,\ps):=\int\, \chd\ps\, d^2x$, which leads to the
condition that the surface integral $i\oint\chd\ga_n\ps\, ds$ vanishes, where
$\ga_n=(\ga,n)=n_\mu \ga_\mu=n\!\!\!\slash\;$ and $n_\mu$ is the outward
oriented normal vectorfield on the boundary.
Imposing local linear boundary conditions which ensure this requirement
amounts to have $\chd\ga_n\ps=0\ $ on the boundary for each pair.
A sufficient condition is to have all modes obeying $\ps=B\ps$ on the boundary,
where the boundary operator $B$ (which is understood to act as the identity in
flavour space) is required to satisfy $B\dag\ga_n B=-\ga_n$ and $B^2=1$.
We shall choose the one-parameter family of boundary operators \cite{HrBa}
\beq
B\equiv B_\th:\equiv i\gaf e^{\th\gaf}\ga_n
\label{hbbc.3}
\eeq
which break the $\gaf$-invariance of the theory, thus making the $N_{\!f}$
flavour theory invariant under $SU(N_{\!f})_V$ instead of $SU(N_{\!f})_L\,
\times SU(N_{\!f})_R$. 
They will be supplemented by suitable boundary conditions for the gauge-field.
These boundary conditions prevent the $U(1)$-current from leaking through the
boundary as they ensure $j\!\cdot\!n=\psd\ga_n\ps=0$ on the boundary. 

For explicit calculations we shall choose the chiral representation
$\ga_0=\sg_1, \ga_1=\sg_2$ and $\gaf=\sg_3$ which implies that
the explicit expressions for the boundary operators at the two ends of
the cylinder (\ref{pro.1}) take the simple form
\beq
B_L=-\left(\begin{array}{cc}0&e^{\th}\\
e^{-\th}&0\end{array}\right)\;({\mbox{at $x^1\!=\!0$}})
\quad\hbox{and}\quad
B_R=+\left(\begin{array}{cc}0&e^{\th}\\
e^{-\th}&0\end{array}\right)\;({\mbox{at $x^1\!=\!L$}})
\quad.
\label{hbbc.6}
\eeq


\subsection{Immediate Consequences for the Spectrum}

The decision to quantize with boundary condition $\ps=B_\th\ps$ with the
boundary operator (\ref{hbbc.3}) has immediate consequences
\cite{WiDu,DuWi,FaGaMuSaSo}:
\begin{itemize}
\item[($i$)]
The Dirac operator has a discrete real spectrum which is {\em asymmetric}
w.r.t. zero.
\item[($ii$)]
The spectrum is empty at zero, i.e. the Dirac operator has {\em no zero
modes}.
\item[($iii$)]
The instanton number $\ q=\frac{e}{4\pi}\int\ep_{\mu\nu}F_{\mu\nu}=
\frac{e}{2\pi}\int E\ \in {\it{\bf R}}\ $ is {\em not quantized}.
\end{itemize}
The first property already indicates that we are not in the situation covered
by the Atiyah-Patodi-Singer-index-theorem.
The second property implies that the generating functional for the fermions
in a given gauge-field background $A$
\beq
Z_F[A,\etd,\et]\;=\;
{1\ov N_F}\int D\psd D\ps\ e^{-\int\psd i\Dsl\ps+i\int\psd\et-i\int\etd\ps}
\label{mfsm.2}
\eeq
is indeed given by the textbook formula
\beq
Z_F[A,\etd,\et]\;=\;
{{\det}_\th(i\Dsl)\ov{\det}_\th(i\pas)}\;\;e^{\,\int\etd(i\Dsl)^{-1}\et}
\label{mfsm.3}
\eeq
and the chiral expectation values follow by taking the logarithmic derivative
\beq
\<\psd(x)P_\pm\ps(x)\>={1\ov Z_F}\;
{\de^2\ov\de\et^{}_\pm(x)\;\de\etd_\pm(x)}\,Z_F\;
\Big\vert_{\et^{}_\pm=\etd_\pm=0}\;.
\label{mfsm.4}
\eeq

Throughout $\th$ is the free parameter in the boundary operator (\ref{hbbc.6}).
The fact that the Feynman-Hellmann-boundary-formula ${d\ov d\th}
\la_k=-\la_k(\ps_k,\gaf\ps_k)$ \cite{O.20} (where the $\la_k$ denote the
eigenvalues of $i\Dsl$) still holds true on the cylinder was the basis for
the analytic determination of the $\th$-dependence of the fermionic
determinant $\det_\th(i\Dsl)$ performed in \cite{DuWi}.


\subsection{Neither Integer nor Fractional but Real Instanton Number}

On a cylinder of finite spatial length the decomposition of a gauge potential
$A_\mu$ is \cite{DuWi}
\beq
\begin{array}{rrr}
eA_0=&-\pa_1\ph+\pa_0\ch&+\frac{2 \pi}{\be}c
\\
eA_1=&+\pa_0\ph+\pa_1\ch&{}
\end{array}
\label{dadt.11}
\eeq
where $\ph$ obeys Dirichlet boundary conditions at the ends ($x^1=0,L)$ and
$\ch$ is a pure gauge degree of freedom which fulfills $\ch(0)+\ch(L)=0$ and
$c\in[-1/2,1/2[$ is the constant harmonic part.
Thus the Dirac operator $i\Dsl=i\ga_\mu(\pa_\mu-ieA_\mu)$ may be factorized
\beq
i\Dsl=G\dag i\Dsl_{\,0}G
\label{dadt.15}
\eeq
where $i\Dsl_{\,0}=\gamma^0(i\pa_0+2\pi c/\beta)+\gamma^1i\pa_1$ is the Dirac
operator with the scalar parts switched off and $G=\mbox{diag}(g^{*-1},g)$
contains the prepotential $g:\equiv e^{-(\ph+i\ch)}$ which is an element of the
complexified gauge-group $U(1)^{*}=S^1\times{\bf R}_{\bf +}$.

On the cylinder there is a one-to-one-correspondence between $\ph$ and
$eF_{01}$ if $\ph$ obeys Dirichlet boundary-conditions at the two ends.
The general field $\ph$ may be decomposed as
\beq
\ph=\sum_{m\geq 0}\sum_{n\geq 1}
\ph_{mn}^+\cos({2\pi mx^0\ov\be})\sin({\pi nx^1\ov L})+
\ph_{mn}^-\sin({2\pi mx^0\ov\be})\sin({\pi nx^1\ov L})
\label{dadt.16}
\eeq
with coefficients $\ph_{mn}^\pm\in{\bf{\rm R}}$ decaying rapidly enough to
make the series converge.
The instanton-number is given by
\beq
q={e\ov 2\pi}\int E\ d^2x={1\ov 2\pi}\int\lap\ph\ d^2x
=-{\be\ov L}\sum_{n\geq 1}{1-(-1)^n\ov2}\,n\,\ph_{0n}^{+}
\label{dadt.17}
\eeq
where the decomposition (\ref{dadt.11}) and the expansion (\ref{dadt.16}) have
been used. Given the result (\ref{dadt.17}) it is obvious that the
instanton-number $q$ may take any real number.


\subsection{Fermionic Propagator w.r.t. Boundary Conditions}

In order to calculate the condensates one needs the Green's function $S_\th$
of the Dirac operator $i\Dsl$ on the cylinder subject to the
XB-boundary-conditions.
In addition to the defining relation $(i\Dsl\;S_\th) (x,y)=\de(x-y)$ this
Green's function obeys the boundary-conditions
\bea
S_\th(x^0\!+\!\be,x^1,y)&=&-\,S_\th(x,y)
\label{fpro.2}\\
(B_L \;S_\th)(x^0,x^1\!=\!0,y)&=&
S_\th(x^0,x^1\!=\!0,y)
\label{fpro.3}\\
(B_R \;S_\th)(x^0,x^1\!=\!L,y)&=&
S_\th(x^0,x^1\!=\!L,y)
\label{fpro.4}
\eea
with $B_{L/R}$ defined in (\ref{hbbc.6}) plus the adjoint relations with
respect to $y$.
The dependence of the gauge-potential has not been made explicit, since from
the factorization-property (\ref{dadt.15}) for the Dirac-operator it follows
at once that $S_\th$ is related to the Green's function $\til S_\th$ of
$i\Dsl_{\,0}$ as
\beq
S_\th(x,y)=G^{-1}(x) \til S_\th(x,y) G^{\dagger\;-1}(y)\ .
\label{fpro.5}
\eeq
Since the field $\ph$ obeys Dirichlet boundary-conditions at the ends
of the cylinder, the deformation matrix $G$ appearing in (\ref{dadt.15}) is
unitary there and the boundary-conditions (\ref{fpro.2}-\ref{fpro.4}) transform
into the identical ones for 
\bdm
\til S_\th(x,y)=
\left(\begin{array}{cc} \til S_{++}& \til S_{+-}\\
\til S_{-+}&\til S_{--}\end{array}\right)
\edm
where the indices refer to chirality.
This Green's function --~which carries the full $c$-dependence of
$S_\th(x,y)$~-- has been determined analytically \cite{DuWi} to read
\bea
\til S_\th(x,y)={i\ov 2\pi}\cdot
\sum_{m,n\in Z\times Z}(-1)^{(m+n)}\cdot
e^{2\pi ic((x^0\!-\!y^0)/\be-n)}
\cdot
\pmatrix{e^\theta/r_{nm}&-(1/s_{nm})\cr
-(1/\bar s_{nm})&e^{-\theta}/\bar r_{nm}},
\label{fpro.6}
\eea
where $r_{nm}=(x^0\!-\!y^0)+i(x^1\!+\!y^1)-(n\beta+2imL)$
and $s_{nm}=(x^0\!-\!y^0)+i(x^1\!-\!y^1)-(n\beta+2imL)$.
From (\ref{fpro.6}) it follows that the
$++$ and $--$ elements at coinciding points inside the cylinder take the forms
\bea
\til S_\th(x,x)_{\pm\!\pm}\!&\!=\!&\!
\pm{e^{\pm\th}\ov4L}\sum_{n\in Z}(-)^n
{e^{\pm2\pi inc}\ov\sin(\pi(\xi-in\tau))}
\nonumber
\\
&\!=\!&\!
\pm{e^{\pm\th}\ov4L}\sum_{n\in Z}(-)^n
{\cos(2\pi nc)\sin(\pi\xi)\rch(\pi n\ta)-
\sin(2\pi nc)\cos(\pi\xi)\rsh(\pi n\ta)\ov
\sin^2(\pi\xi)+\rsh^2(\pi n\ta)}
\label{fpro.8}
\\
\til S_\th(x,x)_{\pm\!\pm}\!&\!=\!&\!
\pm{e^{\pm\th}\ov2\be}\sum_{m\in Z}(-)^m
{e^{-2\pi(m+\xi)c/\tau}\ov\sinh(\pi(m+\xi)/\tau)}
\nonumber
\\
&\!=\!&\!
\pm{e^{\pm\th}\ov2\be}\sum_{m\in Z}(-)^m
{\rch(2\pi(m+\xi)c/\ta)-\rsh(2\pi(m+\xi)c/\ta)\ov
\rsh(\pi(m+\xi)/\ta)}
\label{fpro.10}
\eea
both valid for $c\in[-\ha,\ha]$. The two forms (\ref{fpro.8}, \ref{fpro.10})
are equivalent but enjoy good convergence properties in the two regimes
$\be\gg L$ and $\be\ll L$ respectively.
From  (\ref{fpro.5}) one sees that the chirality violating entries of the
fermionic Green's function lie on the diagonal and take the form
\beq
S_\th(x;x)_{\pm\!\pm}=
e^{\mp2\ph(x)}\til S_\th(x;x)_{\pm\!\pm}
\label{fpro.11}
\eeq
where $\tilde S_{\pm\pm}$ plugged in from (\ref{fpro.8}, \ref{fpro.10})
depends only on the harmonic part $c$ in the decomposition (\ref{dadt.11})
of the gauge-potential.


\subsection{Fermionic Determinant w.r.t. Boundary Conditions}

The arduous step is the computation of the $\th$-dependence of the
fermionic-determinant \cite{DuWi}.
The Dirac-operator and the boundary-conditions are both flavour-neutral.
Thus the determinant is the same for all flavours and it is sufficient to
calculate it for one flavour.
For the explicit calculations we used the gauge-invariant $\ze$-function
definition of the determinant \cite{O.21,ElOdBook}
\beq
\log\det{}_\th(i\Dsl):\equiv{1\ov2}\log\det{}_\th(-\Dsl^2):\equiv
-{1\ov2}{d\ov ds}\bigg\vert_{s=0}\ze{}_\th(-\Dsl^2,s)
\label{fdet.1}
\eeq
and calculated the $\th$-dependence of the $\ze$-function by means of a
boundary-Feynman-Hell\-mann-formula.
Denoting $\{\mu_k\vert k\in {\bf N}\}$ the (positive) eigenvalues of $-\Dsl^2$,
the corresponding $\ze$-function is defined and rewritten as a Mellin-transform
in the usual way
\beq
\ze_\th(s):\equiv\ze_\th(-\Dsl^2,s):\equiv\sum_k \mu_k^{-s}=
{1\ov \Gamma(s)}\int\limits_0^\infty t^{s-1}\tr_\th(e^{-t(-\Dsl^2)})\ dt
\label{fdet.2}
\eeq
for ${\rm Re}(s)>d/2=1$ and its analytic continuation to ${\rm Re}(s)\leq 1$.

The general task was to compute the normalized determinant
\beq
{\det_\th(i\Dsl)\ov\det_0(i\pas)}\equiv
{\det_\th(i\Dsl_{\;1,c})\ov\det_0(i\Dsl_{\;0,0})}
\label{fdet.30}
\eeq
where the first suffix on the r.h.s. indicates whether the scalar part is
switched on and the $c$ refers to the harmonic part in the decomposition
(\ref{dadt.11}).
Together with $\th$ we thus have three parameters to switch off and this leaves
us with $3!=6$ possible choices how to compute the functional determinant
(\ref{fdet.30}) in terms of three factors where each involves one switching
only.
We explicitly followed two of the six choices and found them to agree.

Our first choice was to calculate the functional determinant according to
\beq
{\det_\th(i\Dsl_{\;1,c})\ov\det_0(i\Dsl_{\;0,0})}\ \equiv\
{\det_\th(i\Dsl_{\;1,c})\ov\det_0(i\Dsl_{\;1,c})}\cdot
{\det_0(i\Dsl_{\;1,c})\ov\det_0(i\Dsl_{\;0,c})}\cdot
{\det_0(i\Dsl_{\;0,c})\ov\det_0(i\Dsl_{\;0,0})}
\label{fdet.3}
\eeq
where we got for the first factor the explicit expression
\beq
{\det_\th(i\Dsl_{\;1,c})\ov\det_0(i\Dsl_{\;1,c})}=
\exp\{-{\th\ov 4\pi}\int e\ep_{\mu\nu}F_{\mu\nu}\}=
\exp\{-{\th\ov2\pi}\int\lap\ph\}.
\label{fdet.10}
\eeq
to be multiplied with the result 
\beq
{\det_0(i\Dsl_{\;1,c})\ov\det_0(i\Dsl_{\;0,c})}=\exp\{{1\ov2\pi}\int\ph\lap\ph\}
\label{fdet.11}
\eeq
for the second factor.
It is worth noting that with the first choice the term linear in $\th$ in
the effective action stems from a volume-term (i.e. $a_1(.)$) in the
Seeley-DeWitt-expansion.

Our second choice was to calculate the functional determinant according to
\beq
{\det_\th(i\Dsl_{\;1,c})\ov\det_0(i\Dsl_{\;0,0})}\ \equiv\
{\det_\th(i\Dsl_{\;1,c})\ov\det_\th(i\Dsl_{\;0,c})}\cdot
{\det_\th(i\Dsl_{\;0,c})\ov\det_0(i\Dsl_{\;0,c})}\cdot
{\det_0(i\Dsl_{\;0,c})\ov\det_0(i\Dsl_{\;0,0})}
\label{fdet.12}
\eeq
where we found for the first factor the explicit expression
\beq
{\det_\th(i\Dsl_{\;1,c})\ov\det_\th(i\Dsl_{\;0,c})}=
\exp\{{1\ov2\pi}\int\ph\lap\ph-{\th\ov2\pi}\int\lap\ph\}
\label{fdet.13}
\eeq
to be multiplied with the result 
\beq
{\det_\th(i\Dsl_{\;0,c})\ov\det_0(i\Dsl_{\;0,c})}=\exp\{0\}=1
\label{fdet.14}
\eeq
for the second factor.
It is worth noting that with the second choice the term linear in $\th$ in
the effective action stems from a boundary-term (i.e. $b_1(.)$) in the
Seeley-DeWitt-expansion.

In summary, the scattering of the fermions off the boundary generates a CP-odd
term linear in $\th$ in the effective action for the gauge-bosons which may be
seen as a two-dimensional artificial analogue of the QCD-$\th$-term.

The remaining task (which is to calculate the common third factor in the
factorizations (\ref{fdet.3}) and (\ref{fdet.12}) of the functional determinant)
was addressed by rewriting its logarithm as
\beq
\log{\det_0(i\Dsl_{\;0,c})\ov\det_0(i\Dsl_{\;0,0})}=
-{1\ov2}\int\limits_0^c{d\ov ds}\bigg\vert_{s=0}{d\ov d\til c}\
\ze_0(-\Dsl_{\;0,\til c}^2\;,s)\ d\til c\quad .
\label{fdet.16}
\eeq
and constructing the $s$-derivative at $s=0$ of the $\til c$-derivative of
$\ze_0(-\Dsl_{\;0,\til c}^2\;,s)$ explicitly.
For that aim we computed the heat-kernel of the operator
\bdm
-\Dsl_{\;0,\til c}^{\,2}=-\Big((\pa_0-2\pi i\til c/\be)^2+\pa_1^2
\Big)I_2
\edm
for $\th=0$ on the half-cylinder (see appendix of \cite{DuWi}) and from this
expression we were able to derive 
\beq
\Gamma(c)\equiv-\log{\det_0(i\Dsl_{\;0,c})\ov\det_0(i\Dsl_{\;0,0})}
={V\ov\pi}\sum\nolimits\pri(-1)^{m+n}{\cos(2\pi nc)-1\ov (n\be)^2+(2mL)^2}
\label{fdet.21}
\eeq
as the result for the negative of the logarithm of the last factor of
(\ref{fdet.3}) and (\ref{fdet.12}).
In (\ref{fdet.21}) and in the following the prime in the sum denotes the
omission of the contribution from $m=n=0$.
By performing either the sum over $m$ or the sum over $n$ in (\ref{fdet.21})
one gets
\beq
\Gamma(c)=
\sum\limits_{n\geq1}{(-1)^n\ov n}\
{\mbox{cos}(2n\pi c)-1\ov \mbox{sh}(n\pi\be/2L)}
\label{fdet.23}
\eeq
\beq
\Gamma(c)=
\sum\limits_{m\geq 1}{(-1)^m\ov m}\
{\mbox{ch}(4m\pi cL/\be)-1\ov \mbox{sh}(2m\pi L/\be)}\
+{2\pi L\ov\be}\,c^2
\label{fdet.24}
\eeq
both valid for $c\in[-1/2,1/2]$ and periodically continued otherwise.
These two equivalent forms will be useful in the low- and high- 
temperature expansion of the condensates.


\subsection{Effective Action}

The final step is to combine the classical (euclidean) action of the photon
field, rewritten in the variables (\ref{dadt.11})
\beq
S_B[\ph]\ \equiv\ {1\ov4}F_{\mu\nu}F_{\mu\nu}={1\ov2e^2}\lap\ph\lap\ph\
\label{fdet.25}
\eeq
with the result for the functional determinant (\ref{fdet.30}).
Collecting the contributions (\ref{fdet.10}, \ref{fdet.11}) or
(\ref{fdet.13}, \ref{fdet.14}) as well as (\ref{fdet.23}) or (\ref{fdet.24})
and adding the classical action (\ref{fdet.25}) one ends up with the
effective action (which, of course, does not contain the gauge degree of
freedom $\ch$)
\beq
\Gamma\equiv\Gamma_{\th,N_{\!f}}[c,\ph]\equiv
N_{\!f}\cdot\Gamma(c)+\Gamma_{\th,N_{\!f}}[\ph]
\label{fdet.26}
\eeq
where $\Gamma(c)$ has been given in (\ref{fdet.23}, \ref{fdet.24}) and
$\Gamma_{\th,N_{\!f}}[\ph]$ is
\beq
\Gamma_{\th,N_{\!f}}[\ph]\equiv{1\ov2e^2}\bigg\{
\int\ph\lap^2\ph-\mu^2\!\int\ph\lap\ph+
\mu^2\cdot\th\int\lap\ph\bigg\}
\label{fdet.27}
\eeq
where the fact that the functional determinant is the same for all flavours
has been used.
In (\ref{fdet.27}) $\mu$ is the Schwinger mass (\ref{schwingermass}) which is
the analog of the $\eta^\prime$-mass in 3-flavour-QCD.

In summary, the functional measure takes the form
\beq
d\mu_\th[A]={1\ov Z_\th}\ e^{-\Gamma_{\th,N_{\!f}}[c,\ph]}
\ dc\ D\ph\ \de(\ch)\;D\ch
\label{fdet.28}
\eeq
where we have taken into account that the gauge-variation of the Lorentz
gauge-condition $F:\equiv\pa_\mu A^\mu=\lap\ch$ and the Jacobian of the
transformation from $\{A\}$ to the variables $\{\ph,c,\ch\}$ are independent
of the fields.
Actually, the corresponding determinants cancel each other.

We conclude that the expectation-value of any gauge-invariant operator $O$
(which will not depend on $\ch$) is given by
\beq
\big\langle O \big\rangle=
{\int dc\ D\ph\ \ O\ e^{-\Gamma_{\th,N_{\!f}}[c,\ph]}
\ov \int dc\ D\ph\ \ e^{-\Gamma_{\th,N_{\!f}}[c,\ph]}}
\label{fdet.29}
\eeq
with $\Gamma_{\th,N_{\!f}}[c,\ph]$ given by
(\ref{fdet.26}, \ref{fdet.27}) and (\ref{fdet.23}, \ref{fdet.24}) .


\section{Condensates in a Finite Volume}


The general result (\ref{fdet.29}) may be applied to calculate the chiral
condensates
\bea
\<\ps\dag(x)P_\pm\ps(x)\>=
{\int dc\, D\ph\ \ S_\th(x,x)_{\pm\!\pm}\
e^{-\Gamma_{\th,N_{\!f}}[c,\ph]}
\ov
\int dc\, D\ph\ \
e^{-\Gamma_{\th,N_{\!f}}[c,\ph]}}
\label{rtpc.1}
\eea
with $S_\th$ from (\ref{fpro.11}) and $\Gamma_\th$ from (\ref{fdet.26}).
Both the (exponentiated) action and the Green's function factorize 
into parts which only depend on $c$ and $\ph$, respectively.
Thus (\ref{rtpc.1}) factorizes as
\beq
\<\ps\dag(x)P_\pm\ps(x)\>=C^{\pm}(x)\cdot D^{\pm}(x)
\label{rtpc.2}
\eeq
with $x^0$-independent factors
\bea
C^{\pm}(x^1)&=&
{\int dc\ \til S_\th(x,x)_{\pm\!\pm}\ e^{-N_{\!f}\Gamma(c)}
\ov\int dc\ \ e^{-N_{\!f}\Gamma(c)}}\quad,
\label{rtpc.3}
\\
D^{\pm}(x^1)&=&
{\int\ D\ph\ e^{\mp 2\ph(x)-\Gamma_{\th,N_{\!f}}[\ph]}
\ov\int\ D\ph\ \ e^{-\Gamma_{\th,N_{\!f}}[\ph]}}
\label{rtpc.4}
\eea
which depend on the parameters $\th,N_{\!f},\be,L$.
Here and below the $c$-integrals extend over the period $[-1/2,1/2]$, whereas
the field $\ph$ is subject to Dirichlet boundary conditions at the two ends
$x^1\!=\!0$ and $x^1\!=\!L$.
The next step is to evaluate the factors $C^{\pm}$ and $D^{\pm}$ in
(\ref{rtpc.2}) for given circumference $\be$ and length $L$ of the cylinder
and given values of $\th$ and $N_{\!f}$.


\subsection{Harmonic Integral}

In order to evaluate the first factor in (\ref{rtpc.2}) it is worth noticing
that the two forms (\ref{fdet.23}, \ref{fdet.24}) allow one to write the factor
$\exp\{-\Gamma(c)\}$ in the two equivalent versions
\beq
e^{-\Gamma(c)}=
{\theta_3(c,i\tau)\ov\theta_3(0,i\tau)}
\label{thet.1}
\eeq
\beq
e^{-\Gamma(c)}=
e^{-\pi c^2/\tau}\;{\theta_3(ic/\tau,i/\tau)\ov\theta_3(0,i/\tau)}\;.
\label{thet.2}
\eeq
Here we employed the notation
\beq
\theta_3(u,\om)=\sum_{n\in Z}e^{2\pi inu}q^{n^2}=
1+2\sum_{n\geq1}\cos(2n\pi u)q^{n^2}\qquad\quad(q\equiv e^{i\pi\om})
\label{thet.4}
\eeq
for the parameters $\om=i\ta$ and $\om=i/\ta$ (giving real nome $q\in]0,1[$)
respectively.
In deriving (\ref{thet.1}) and (\ref{thet.2}) we used the infinite-product
expansion \cite{GrRy}
\beq
\theta_3(u,\om)=\prod_{n\geq1}
(1-q^{2n})(1+2q^{2n-1}\cos(2\pi u)+q^{2(2n-1)})
\label{thet.5}
\eeq
and the addition theorem \cite{AbSt}
\beq
\ln\Big({\theta_3(u\!+\!v,\om)\ov\theta_3(u\!-\!v,\om)}\Big)=
4\sum\limits_{n\geq1}{(-1)^n\ov n}{q^n\ov 1-q^{2n}}
\sin(2\pi nu)\sin(2\pi nv)
\label{thet.6}
\eeq
respectively.
The Poisson resummation lemma makes sure that
\beq
\sum_{n\in Z} e^{-(x+n)^2\cdot t}=
\sqrt{\pi/t}\sum_{n\in Z} e^{-\pi^2n^2/t}e^{2\pi inx}
\qquad\qquad(x\in{\bf{\it R}},t>0)
\label{pois.0}
\eeq
from which we derive the theta-function duality relation
\beq
\sqrt{\ta}\;\theta_3(c,i\ta)=e^{-\pi c^2/\ta}\;\theta_3(\pm ic/\ta,i/\ta)
\qquad\qquad(c\in{\bf{\it R}},\ta>0)
\label{thet.3}
\eeq
which in turn allows to confirm the identity of the right-hand sides of
(\ref{thet.1}) and (\ref{thet.2}).

Starting from expression (\ref{rtpc.3}) and plugging in the Green's function
(\ref{fpro.8}) as well as (\ref{thet.1}) or alternatively the Green's function
(\ref{fpro.10}) along with (\ref{thet.2}) one ends up with
\bea
C^{\pm}(x^1)\!&\!=\!&\!\pm{e^{\pm\th}\ov4L}\sum_{n\in Z}
(-1)^n\,
{\sin(\pi\xi)\rch(\pi n\ta)\ov\sin^2(\pi\xi)+\rsh^2(\pi n\ta)}\cdot
{\int\cos(2\pi nc)\,\theta_3^{N_{\!f}}(c,i\tau)\ dc\ov
\int\,\theta_3^{N_{\!f}}(c,i\tau)\ dc}
\label{hilf.1}
\\
\nonumber
\\
C^{\pm}(x^1)\!&\!=\!&\!\pm{e^{\pm\th}\ov2\be}\sum_{m\in Z}
(-1)^m\,
{1\ov\rsh(\pi(m\!+\!\xi)/\ta)}\cdot
{\int\rch(2\pi(m\!+\!\xi)c/\tau)\,
e^{-N_{\!f}\pi c^2/\ta}\;\theta_3^{N_{\!f}}(ic/\tau,i/\tau)\ dc\ov
\int\,e^{-N_{\!f}\pi c^2/\tau}\;\theta_3^{N_{\!f}}(ic/\tau,i/\tau)\ dc}
\nonumber
\\
&{}&
\label{hilf.2}
\eea
which is independent of $x^0$ as required by translation-invariance.
In the last step we have taken advantage from the fact that $\theta_3(.,.)$
is symmetric in its first argument and the $c$-integration is from $-1/2$ to
$1/2$.


\subsection{Scalar Integral}

For the evaluation of the second factor in (\ref{rtpc.2}) we recall that the
integration extends over fields $\ph$ which are periodic in the $x^0$ and
satisfy Dirichlet boundary-conditions at $x^1=0,L$.
The first thing to do is to perform the gaussian integrals to get ($\lap'$ is
the Laplacian taking derivatives w.r.t. $x'$)
\beq
D^{\pm}(x^1)\;=\;\exp\Big\{{2\pi\ov N_{\!f}}K(x,x)\Big\}
\cdot
\exp\Big\{\pm{\th\ov2}\cdot\!\int\lap'K(x,x')\ d^2x'
\pm{\th\ov2}\cdot\!\int\lap'K(x',x)\ d^2x'\Big\}
\label{rtpc.7}
\eeq
where the integration is over $x'$ and the kernel (employing the Schwinger mass
(\ref{schwingermass}))
\beq
K(x,y)=
\<x\vert{\mu^2\ov-\lap(-\lap+\mu^2)}\vert y\>=
\<x\vert{1\ov-\lap}\vert y\>-\< x\vert{1\ov-\lap+\mu^2}\vert y\>
\label{rtpc.8}
\eeq
is with respect to Dirichlet boundary conditions.
From its explicit form one finds \cite{DuWi}
\bea
K(x,x)&=&{1\ov 2\pi}\sum_{n\geq1}\Big(1-\cos(2n\pi\xi)\Big)
\Big({{\rm cth}(n\pi\ta)\ov n}-
(n\rightarrow\sqrt{n^2\!+\!(\mu L/\pi)^2\,})\Big)
\label{rtpc.10}
\\
\nonumber
\\
K(x,x)&=&{\xi(1-\xi)\ov 2\ta}+
{\rch(\mu L(1-2\xi))-\rch(\mu L)\ov 2\mu\be\rsh(\mu L)}+
\nonumber
\\
{}&{}&
{1\ov 2\pi}\sum_{m\geq1}
{\rch(m\pi/\ta)-\rch(m\pi(1-2\xi)/\ta)\ov m\rsh(m\pi/\ta)}-
(m\rightarrow\sqrt{m^2\!+\!(\mu\be/2\pi)^2\,})
\label{rtpc.11}
\eea
which is perfectly finite as well as
\beq
\int\lap'K(x,x')\ d^2x'=\int\lap'K(x',x)\ d^2x'=
{\rsh(\mu L(1-\xi))+\rsh(\mu L\xi)\ov\rsh(\mu L)}-1
\label{rtpc.12}
\eeq
from which the factor (\ref{rtpc.7}) can be computed.
As one can see, all three expressions (\ref{rtpc.10}, \ref{rtpc.11},
\ref{rtpc.12}) tend to zero as $x$ approaches the boundary.
The result for (\ref{rtpc.7}) is then found to read
\bea
D^{\pm}(x^1)&=&
\exp\Big\{{1\ov N_{\!f}}\sum\limits_{n\geq1}
\Big(1-\cos(2n\pi\xi)\Big)
\Big({{\rm cth}(n\pi\tau)\ov n}-
(n\rightarrow\sqrt{n^2\!+\!(\mu L/\pi)^2\,})\Big)\Big\}
\cdot
\nonumber
\\
&{}&
\exp\Big\{\pm\th\cdot\Big({\rsh(\mu L(1\!-\!\xi))+\rsh(\mu L\xi)
\ov\rsh(\mu L)}-1\Big)\Big\}
\label{rtpc.15}
\\
\nonumber
\\
D^{\pm}(x^1)&=&
\exp\Big\{{2\pi\ov N_{\!f}}
\Big({\xi(1\!-\!\xi)\ov 2\tau}+{\rch(\mu L(1\!-\!2\xi))-\rch(\mu L)
\ov 2\mu\be\ \rsh(\mu L)}\Big)\Big\}
\cdot
\nonumber
\\
&{}&
\exp\Big\{{1\ov N_{\!f}}\sum\limits_{m\geq1}
{\rch(m\pi/\tau)-\rch(m\pi(1\!-\!2\xi)/\tau)
\ov m\ \rsh(m\pi/\tau)}-
(m\rightarrow\sqrt{m^2\!+\!(\mu\be/2\pi)^2\,})\Big\}
\cdot
\nonumber
\\
&{}&
\exp\Big\{\pm\th\cdot\Big({\rsh(\mu L(1\!-\!\xi))+\rsh(\mu L\xi)
\ov\rsh(\mu L)}-1\Big)\Big\}
\label{rtpc.16}
\eea
which is independent of $x^0$ as required by translation-invariance.


\subsection{Analytical Expression for Condensate at Arbitrary Points}

The chiral condensate (\ref{rtpc.2}) is simply the product of
(\ref{hilf.1}) and (\ref{rtpc.15}) or equivalently
(\ref{hilf.2}) and (\ref{rtpc.16}).
Rescaled by natural units (\ref{natuni}), it takes the form
\bea
{\<\psd P_{\pm}\ps\>(x^1)\ov(|e|/\sqrt{\pi})}\!&\!=\!&\!
\pm{e^{\pm\th\cdot\rch(\la\sqrt{N_{\!f}}(1-2\xi)/2)/\rch(\la\sqrt{N_{\!f}}/2)}
\ov4\la}\cdot
\nonumber
\\
&{}&\!
\sum_{n\in Z}(-1)^n\,
{\sin(\pi\xi)\rch(\pi n\ta)\ov\sin^2(\pi\xi)+\rsh^2(\pi n\ta)}\cdot
{\int\cos(2\pi nc)\,\theta_3^{N_{\!f}}(c,i\tau)\ dc\ov
\int\,\theta_3^{N_{\!f}}(c,i\tau)\ dc}\cdot
\nonumber
\\
&{}&\!
\exp\Big\{{1\ov N_{\!f}}\sum\limits_{n\geq1}
\Big(1-\cos(2n\pi\xi)\Big)
\Big({{\rm cth}(n\pi\tau)\ov n}-
(n\rightarrow\sqrt{n^2\!+N_{\!f}(\la/\pi)^2\,})\Big)\Big\}
\label{copc.1}
\\
\nonumber
\\
{\<\psd P_{\pm}\ps\>(x^1)\ov(|e|/\sqrt{\pi})}\!&\!=\!&\!
\pm{e^{\pm\th\cdot\rch(\la\sqrt{N_{\!f}}(1-2\xi)/2)/\rch(\la\sqrt{N_{\!f}}/2)}
\ov2\sg}\cdot
\nonumber
\\
&{}&\!
\sum_{m\in Z}(-1)^m\,
{1\ov\rsh(\pi(m\!+\!\xi)/\ta)}\cdot
{\int\rch(2\pi(m\!+\!\xi)c/\ta)\,
e^{-N_{\!f}\pi c^2/\ta}\;\theta_3^{N_{\!f}}(ic/\tau,i/\tau)\;dc\ov
\int\,e^{-N_{\!f}\pi c^2/\tau}\;\theta_3^{N_{\!f}}(ic/\tau,i/\tau)\;dc}\!\cdot
\nonumber
\\
&{}&\!
\exp\Big\{{\pi\ov N_{\!f}}
\Big({\xi(1\!-\!\xi)\ov\tau}+
{\rch(\la\sqrt{N_{\!f}}(1\!-\!2\xi))-\rch(\la\sqrt{N_{\!f}})
\ov\sg\sqrt{N_{\!f}}\ \rsh(\la\sqrt{N_{\!f}})}\Big)\Big\}\cdot
\label{copc.2}
\\
&{}&\!
\exp\Big\{{1\ov N_{\!f}}\sum\limits_{m\geq1}
{\rch(m\pi/\tau)-\rch(m\pi(1\!-\!2\xi)/\tau)
\ov m\ \rsh(m\pi/\tau)}-
(m\rightarrow\sqrt{m^2\!+N_{\!f}(\sg/2\pi)^2\,})\Big\}
\nonumber
\eea
where the $\xi$-independent but $\th$-dependent parts of the two factors have
cancelled.
Note that the two forms (\ref{copc.1}) and (\ref{copc.2}) are identical for
any finite (dimensionless) lengths $\sg$ and $\la$, but enjoy excellent
convergence properties in the low- ($\ta\gg 1$) and high- ($\ta\ll 1$)
temperature regimes, respectively.


\section{Numerical Evaluation for $N_{\!f}\!=\!1$ and $N_{\!f}\!=\!2$}


The general result (\ref{copc.1}, \ref{copc.2}) shall be specialized to the
cases $N_f\!=\!1$ and $N_f\!=\!2$ as the integration over $c$ can be done in
these cases.
The aim is to collect specific observations concerning the difference between
the single-flavour and multi-flavour case.


\subsection{Specialization to $N_{\!f}=1$ at Arbitrary Points}

For one flavour the $c$-integrals in (\ref{copc.1}, \ref{copc.2}) can be
performed. Taking into account the fact that they extend over the
interval [-1/2,1/2] the result is found to read
\bea
{\<\psd P_{\pm}\ps\>\ov(|e|/\sqrt{\pi})}\!\!&\!=\!&\!
\pm{e^{\pm\th\cdot\rch(\la/2-\la\xi)/\rch(\la/2)}\ov4\la}
\sum_{n\in Z}(-1)^n\,
{\sin(\pi\xi)\rch(\pi n\ta)\ov\sin^2(\pi\xi)+\rsh^2(\pi n\ta)}\;
\exp(-n^2\pi\ta)\cdot
\nonumber
\\
&{}&\!
\exp\Big\{\sum\limits_{n\geq1}
\Big(1-\cos(2n\pi\xi)\Big)
\Big({{\rm cth}(n\pi\tau)\ov n}-
(n\rightarrow\sqrt{n^2\!+\!(\la/\pi)^2\,})\Big)\Big\}
\label{copc.5}
\\
\nonumber
\\
{\<\psd P_{\pm}\ps\>\ov(|e|/\sqrt{\pi})}\!\!&\!=\!&\!
\pm{e^{\pm\th\cdot\rch(\la/2-\la\xi)/\rch(\la/2)}\ov2\sg}
\sum_{m\in Z}(-1)^m\,
{1\ov\rsh(\pi(m+\xi)/\ta)}\times
\nonumber
\\
&{}&\!
{\sum\limits_{k\in Z}
e^{\pi(m+\xi)(2k+m+\xi)/\ta}
({\rm erf}({(k+m+\xi+1/2)\sqrt{\pi}\ov\sqrt{\ta}})
\!-\!{\rm erf}({(k+m+\xi-1/2)\sqrt{\pi}\ov\sqrt{\ta}}))
\ov2
}\cdot
\nonumber
\\
&{}&\!
\exp\Big\{\pi
\Big({\xi(1\!-\!\xi)\ov\tau}+{\rch(\la(1\!-\!2\xi))-\rch(\la)
\ov\sg\ \rsh(\la)}\Big)\Big\}\cdot
\nonumber
\\
&{}&\!
\exp\Big\{\sum\limits_{m\geq1}
{\rch(m\pi/\tau)-\rch(m\pi(1\!-\!2\xi)/\tau)
\ov m\ \rsh(m\pi/\tau)}-
(m\rightarrow\sqrt{m^2\!+\!(\sg/2\pi)^2\,})\Big\}\qquad
\label{copc.6}
\eea
where again the two equivalent representations (\ref{copc.5}) and (\ref{copc.6})
enjoy excellent convergence properties in the low- ($\ta\gg 1$) and high-
($\ta\ll 1$) temperature regimes, respectively.


\subsection{Specialization to $N_{\!f}=2$ at Arbitrary Points}

For two flavours the $c$-integrals in (\ref{copc.1}, \ref{copc.2}) can be
performed. Taking into account the fact that they extend over the
interval [-1/2,1/2] the result is found to read
\bea
{\<\psd P_{\pm}\ps\>\ov(|e|/\sqrt{\pi})}\!&\!=\!&\!
\pm{e^{\pm\th\cdot\rch(\la/\sqrt{2}-\sqrt{2}\la\xi)/\rch(\la/\sqrt{2})}\ov4\la}
\sum_{n\in Z}(-1)^n\,
{\sin(\pi\xi)\rch(\pi n\ta)\ov\sin^2(\pi\xi)+\rsh^2(\pi n\ta)}
{\sum\limits_{k\in Z}e^{-(k^2\!+\!(n\!-\!k)^2)\pi\ta}\ov
\sum\limits_{k\in Z}e^{-2k^2\pi\ta}}\cdot
\nonumber
\\
&{}&\!
\exp\Big\{{1\ov 2}\sum\limits_{n\geq1}
\Big(1-\cos(2n\pi\xi)\Big)
\Big({{\rm cth}(n\pi\tau)\ov n}-
(n\rightarrow\sqrt{n^2\!+\!2(\la/\pi)^2\,})\Big)\Big\}
\label{neoc.5}
\\
\nonumber
\\
{\<\psd P_{\pm}\ps\>\ov(|e|/\sqrt{\pi})}\!&\!=\!&\!
\pm{e^{\pm\th\cdot\rch(\la/\sqrt{2}-\sqrt{2}\la\xi)/\rch(\la/\sqrt{2})}\ov2\sg}
\sum_{m\in Z}(-1)^m\,
{1\ov\rsh(\pi(m+\xi)/\ta)}\times
\nonumber
\\
&{}&\!
{\sum\limits_{q\in Z}\sum\limits_{p\in Z}
{1+(-1)^{p+q}\ov2}
e^{\pi((p+m+\xi)^2-p^2-q^2)/2\ta}
({\rm erf}({(p+m+\xi+1)\sqrt{\pi}\ov\sqrt{2\ta}})
-{\rm erf}({(p+m+\xi-1)\sqrt{\pi}\ov\sqrt{2\ta}}))
\ov
2\sum\limits_{q\in Z}e^{-\pi q^2/2\ta}
}\!\cdot\!
\nonumber
\\
&{}&\!
\exp\Big\{{\pi\ov2}
\Big({\xi(1\!-\!\xi)\ov\tau}+{\rch(\sqrt{2}\la(1\!-\!2\xi))-\rch(\sqrt{2}\la)
\ov\sqrt{2}\sg\ \rsh(\sqrt{2}\la)}\Big)\Big\}\cdot
\nonumber
\\
&{}&\!
\exp\Big\{{1\ov 2}\sum\limits_{m\geq1}
{\rch(m\pi/\tau)-\rch(m\pi(1\!-\!2\xi)/\tau)
\ov m\ \rsh(m\pi/\tau)}-
(m\rightarrow\sqrt{m^2\!+\!2(\sg/2\pi)^2\,})\Big\}\qquad
\label{neoc.6}
\eea
where again the two equivalent representations (\ref{neoc.5}) and (\ref{neoc.6})
enjoy excellent convergence properties in the low- ($\ta\gg 1$) and high-
($\ta\ll 1$) temperature regimes, respectively.


\subsection{Numerical Evaluation of $\xi$-Dependence}

\begin{figure}
\begin{tabular}{lll}
\epsfig{file=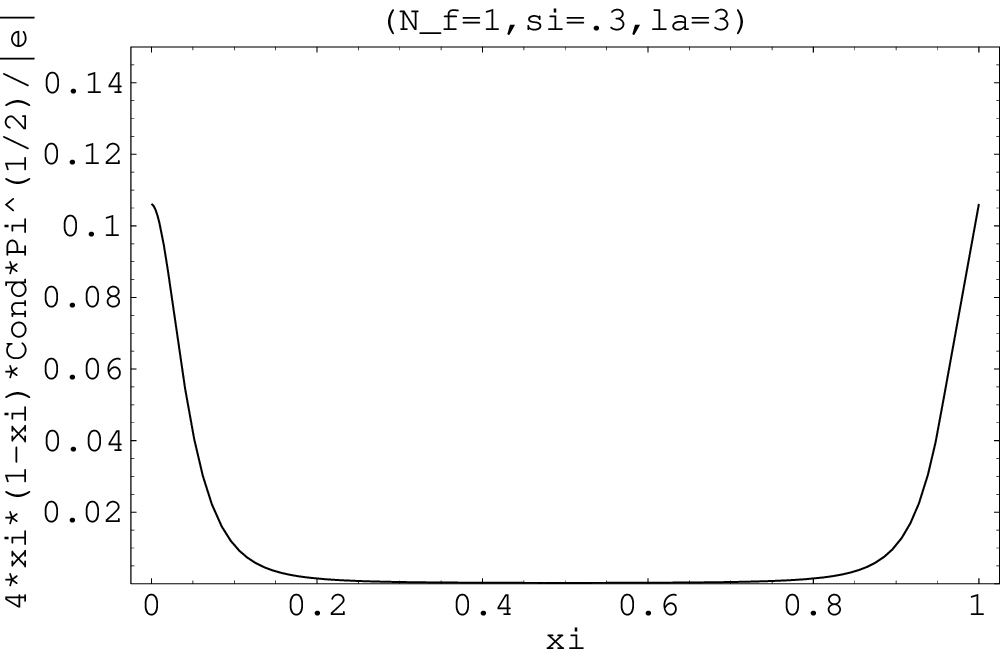,width=6.7cm,angle=90}&
\epsfig{file=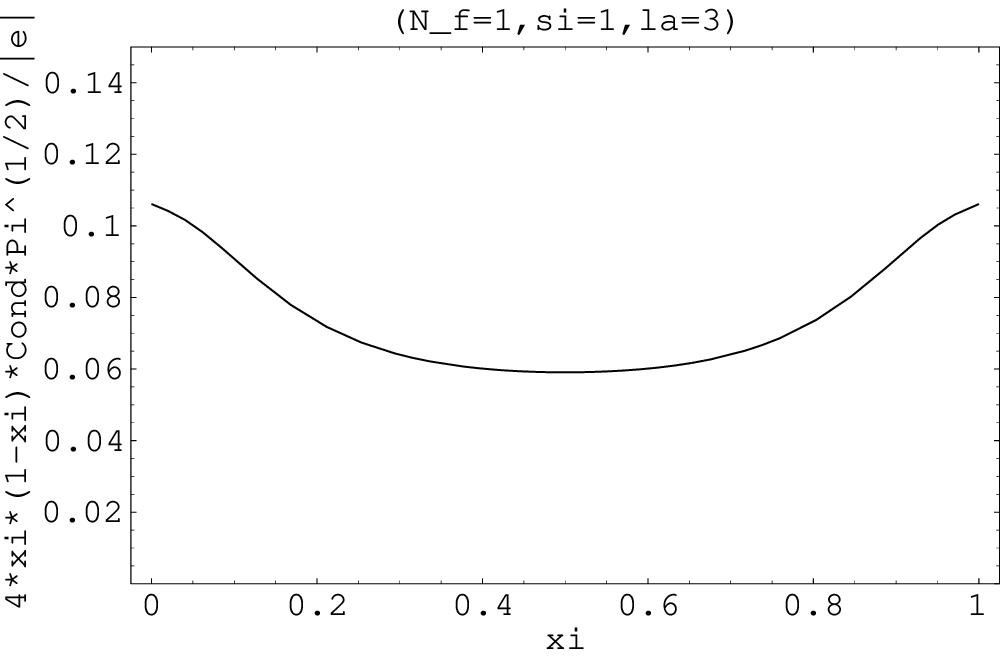,width=6.7cm,angle=90}&
\epsfig{file=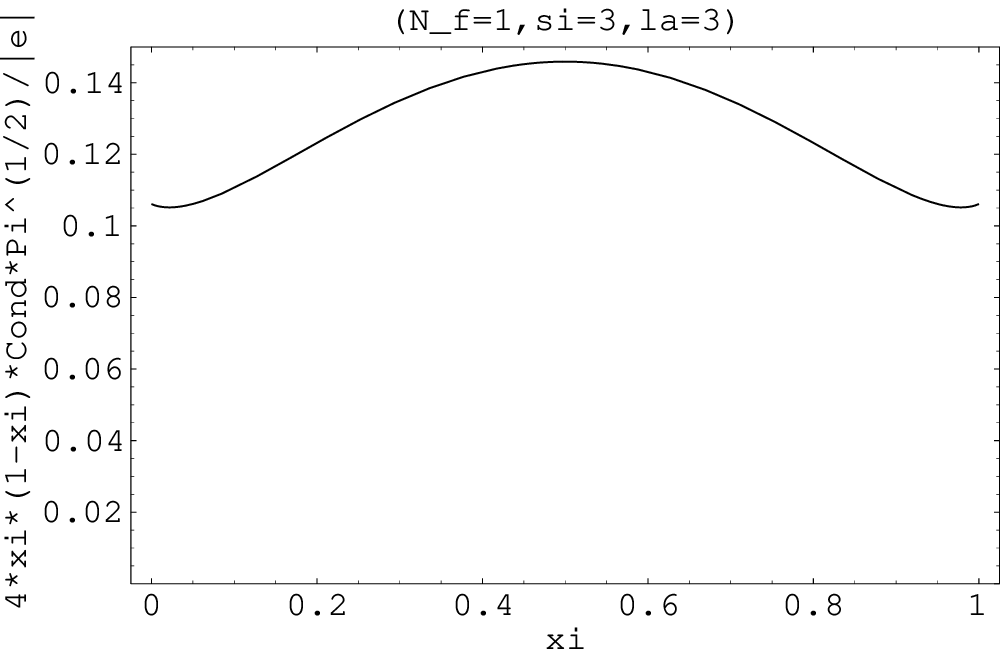,width=6.7cm,angle=90}\\
\epsfig{file=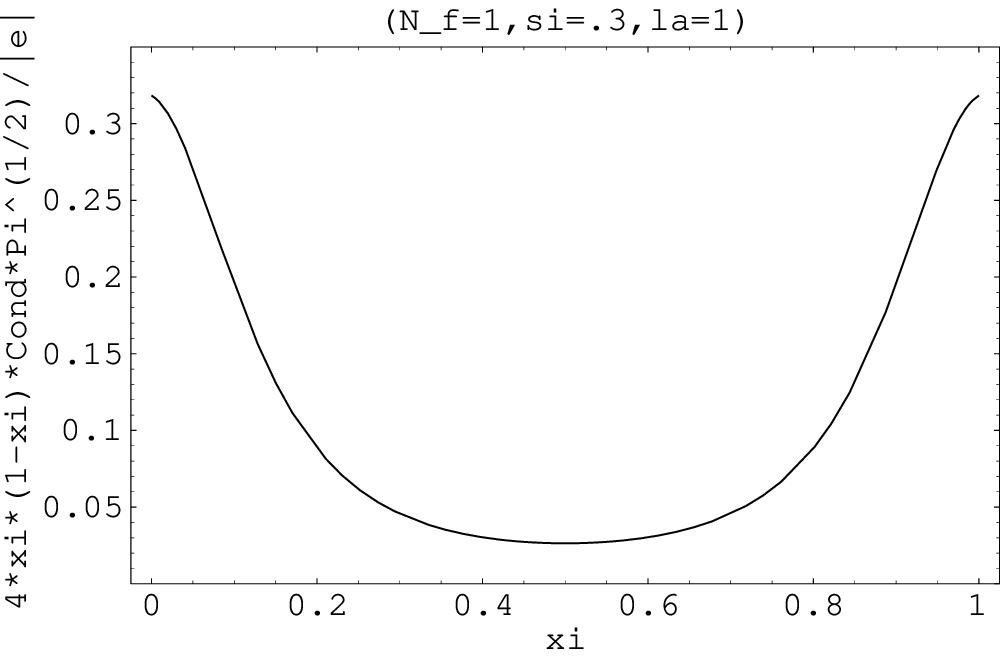,width=6.7cm,angle=90}&
\epsfig{file=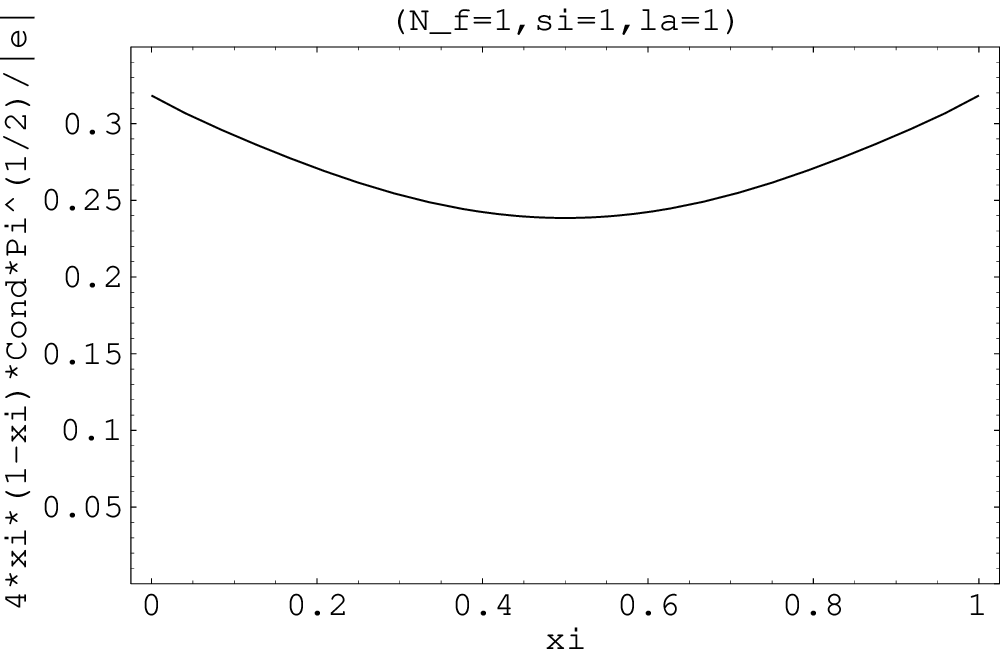,width=6.7cm,angle=90}&
\epsfig{file=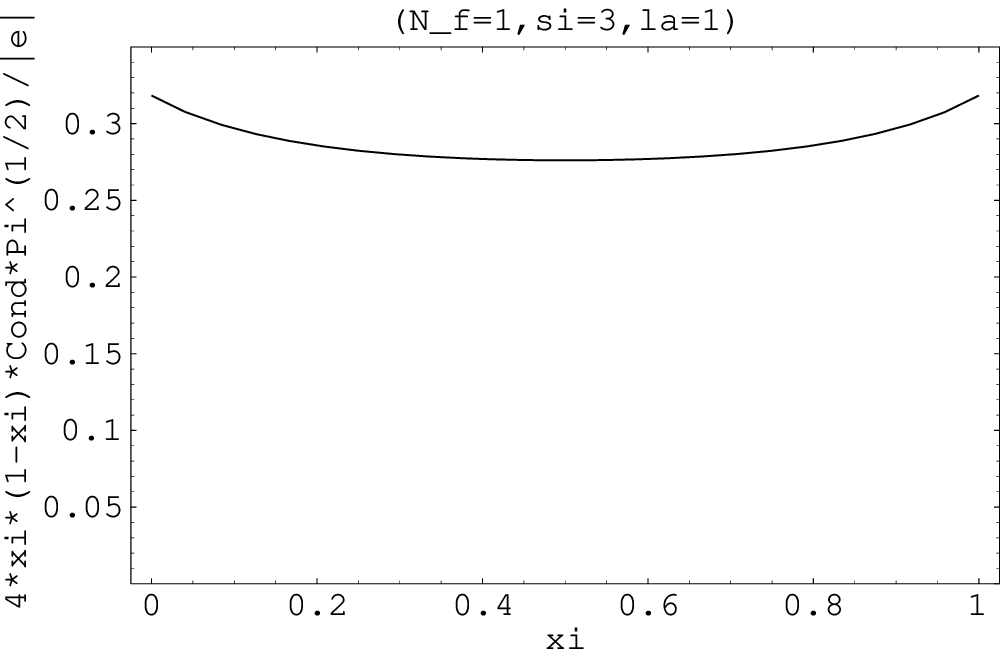,width=6.7cm,angle=90}\\
\epsfig{file=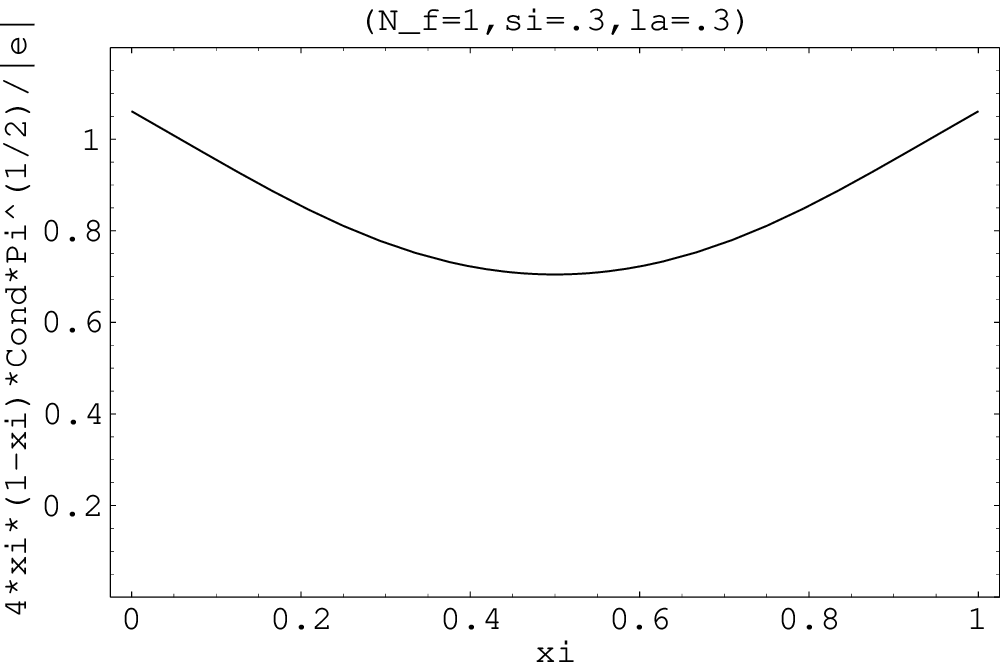,width=6.7cm,angle=90}&
\epsfig{file=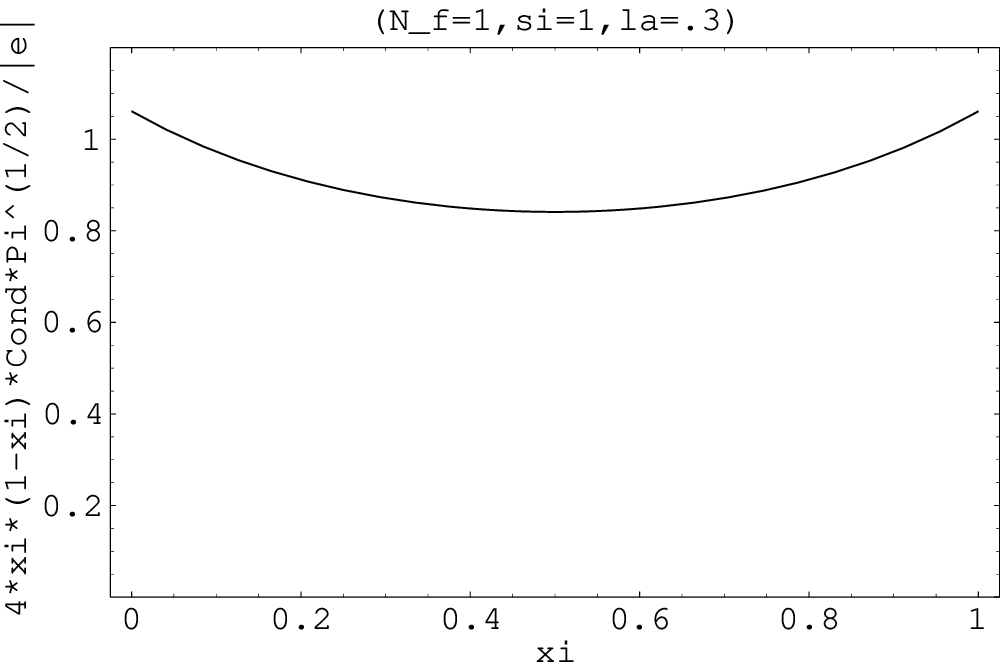,width=6.7cm,angle=90}&
\epsfig{file=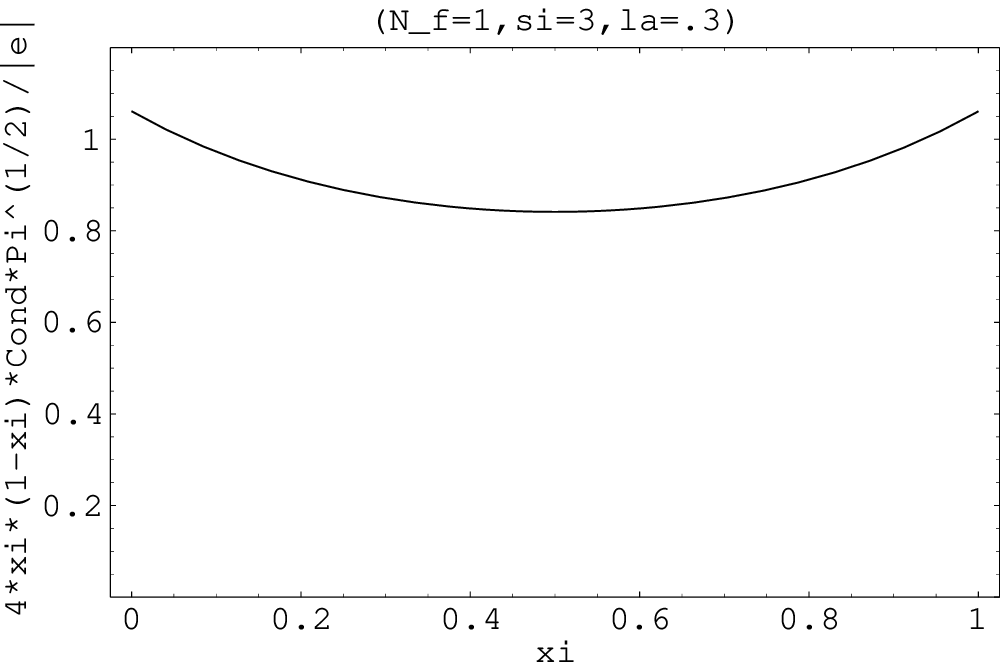,width=6.7cm,angle=90}
\end{tabular}
\caption{Spatial Dependence of $4\xi(1\!\!-\!\xi)|\<\psi\dag P_\pm\psi\>|/
\mu_1$ for $N_{\!f}=1$ ($\xi\!=\!x^1\!/L$).}
\label{fig2}
\end{figure}

\begin{figure}
\begin{tabular}{lll}
\epsfig{file=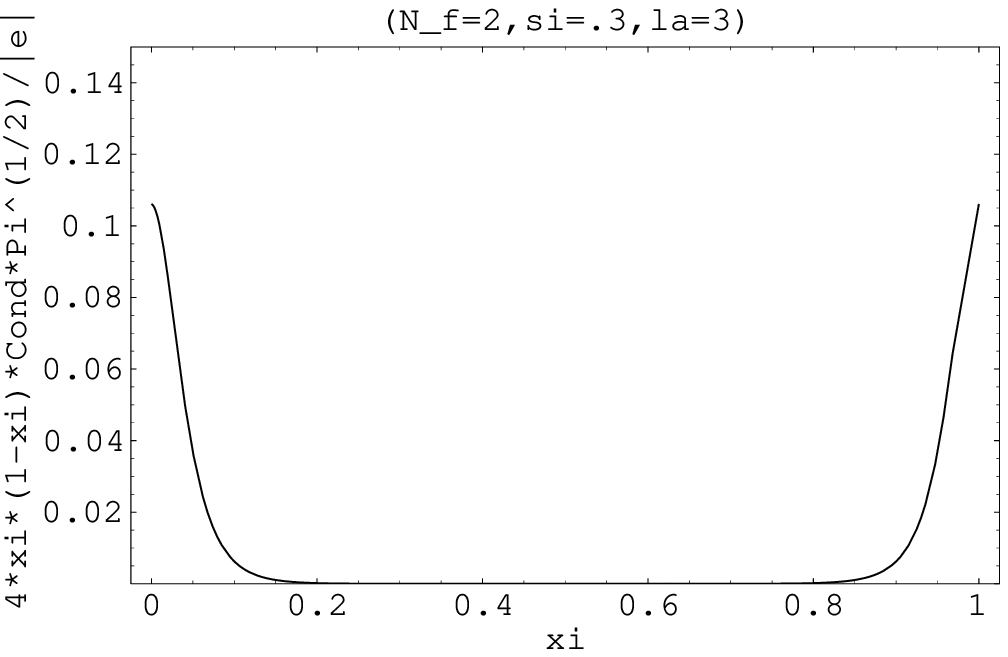,width=6.7cm,angle=90}&
\epsfig{file=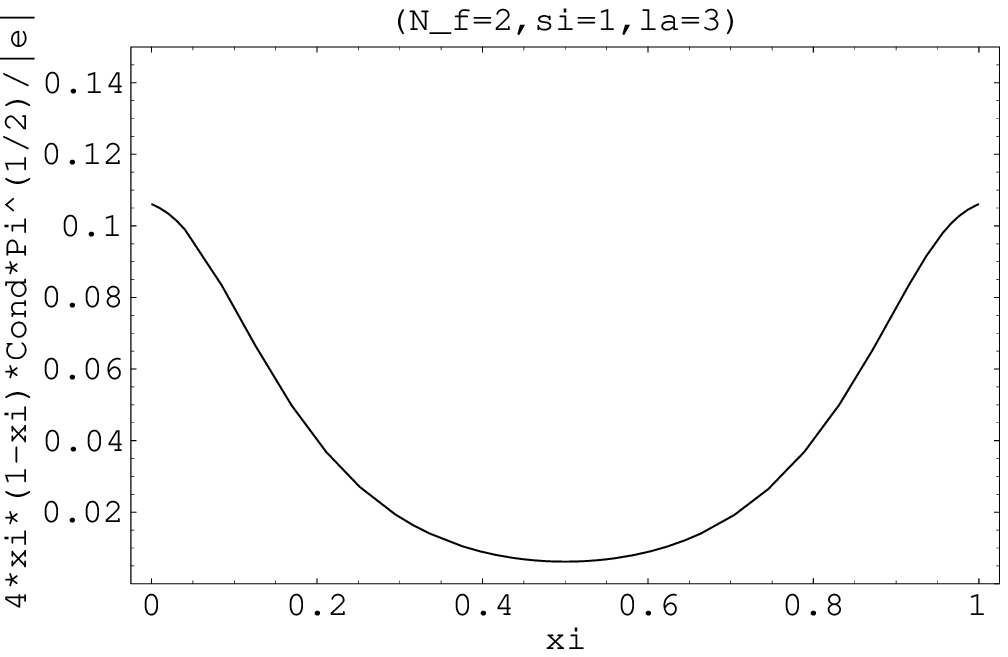,width=6.7cm,angle=90}&
\epsfig{file=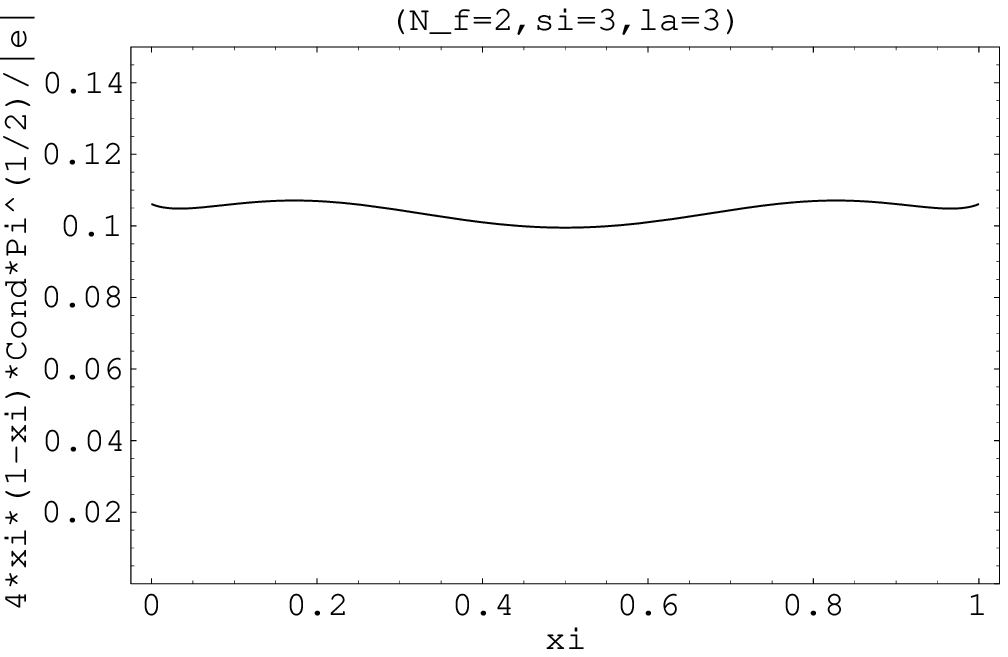,width=6.7cm,angle=90}\\
\epsfig{file=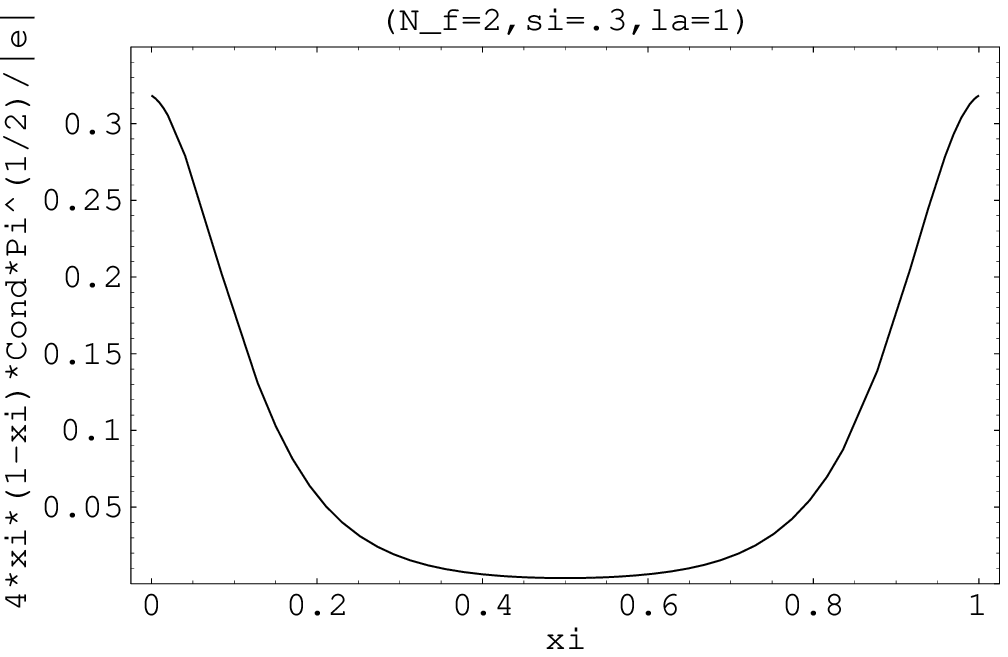,width=6.7cm,angle=90}&
\epsfig{file=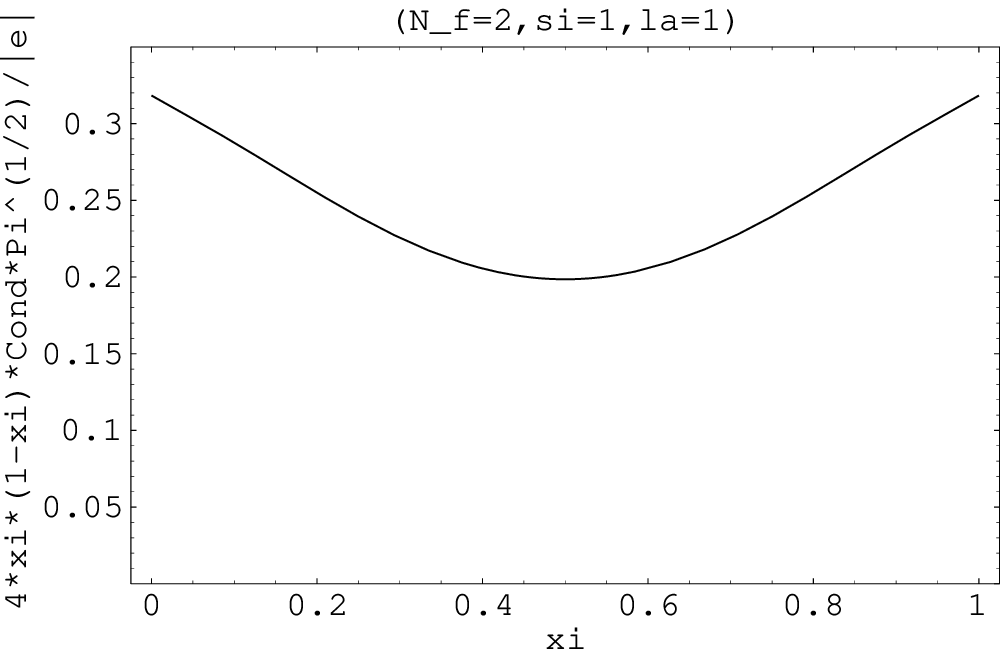,width=6.7cm,angle=90}&
\epsfig{file=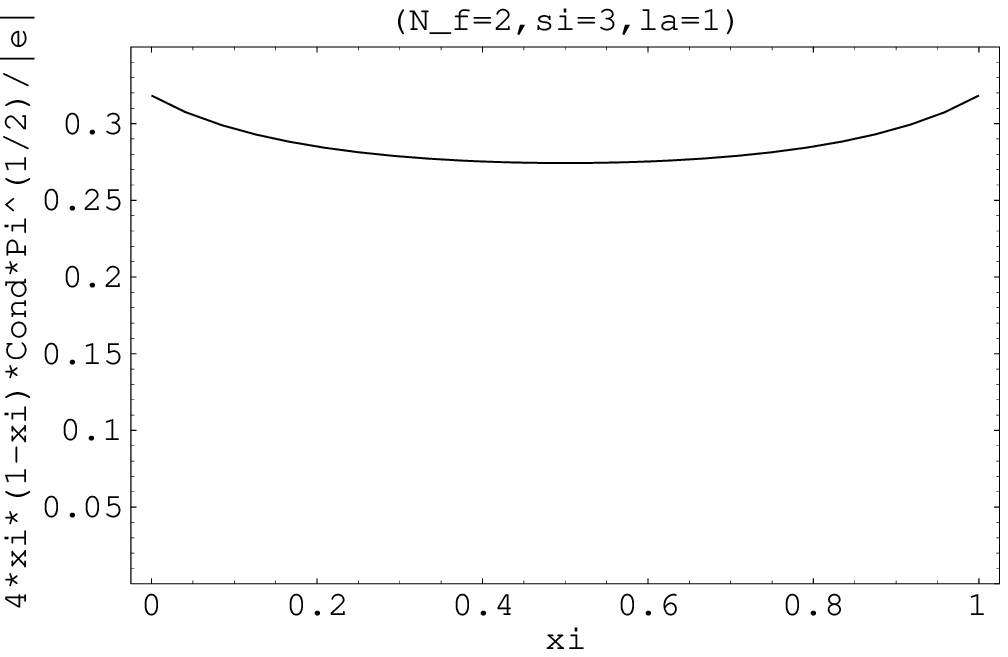,width=6.7cm,angle=90}\\
\epsfig{file=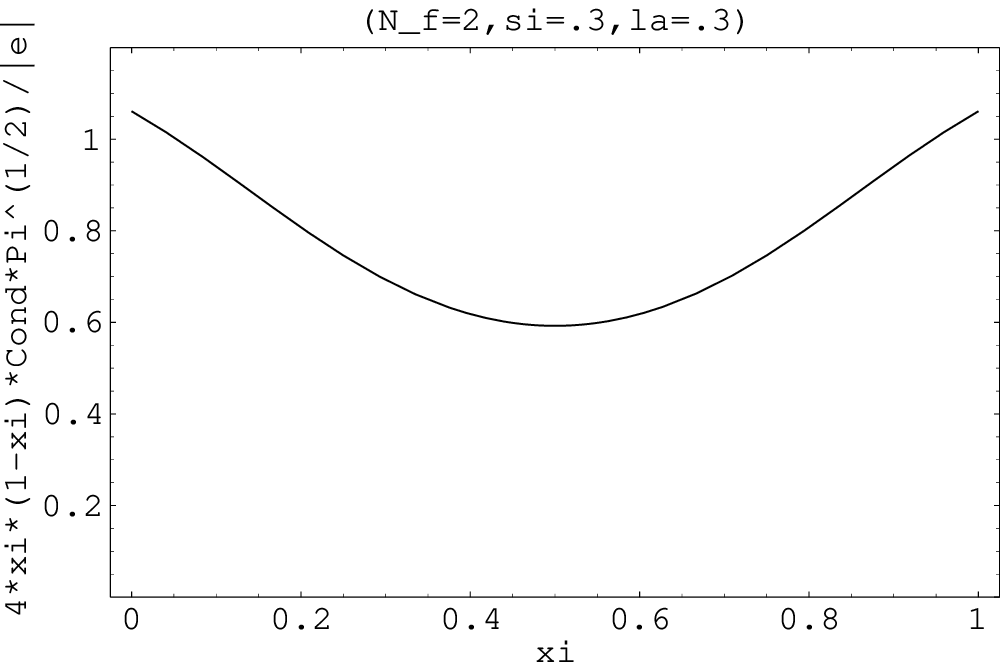,width=6.7cm,angle=90}&
\epsfig{file=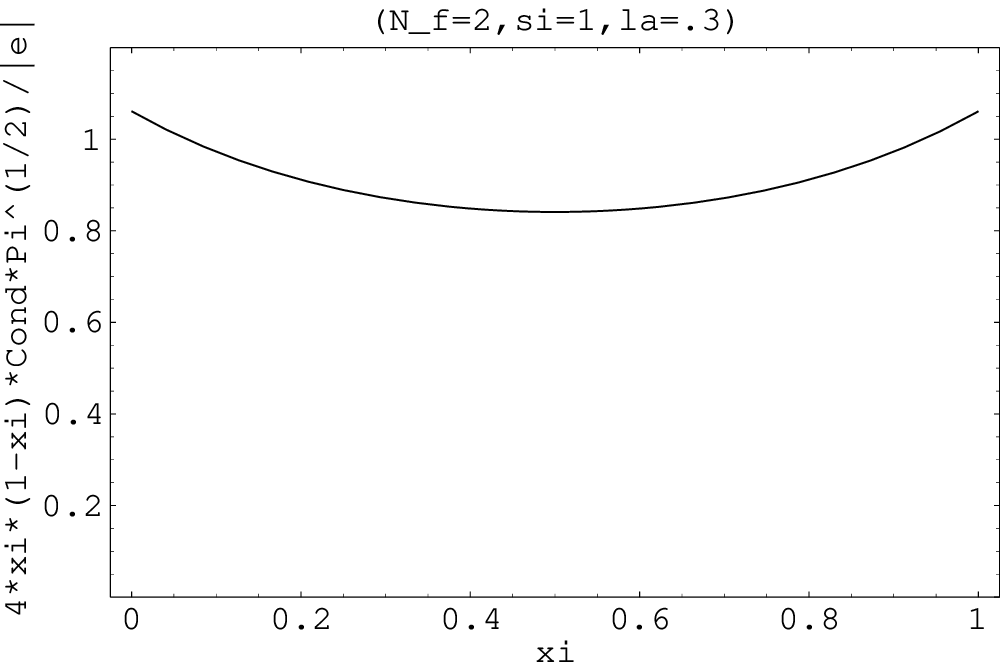,width=6.7cm,angle=90}&
\epsfig{file=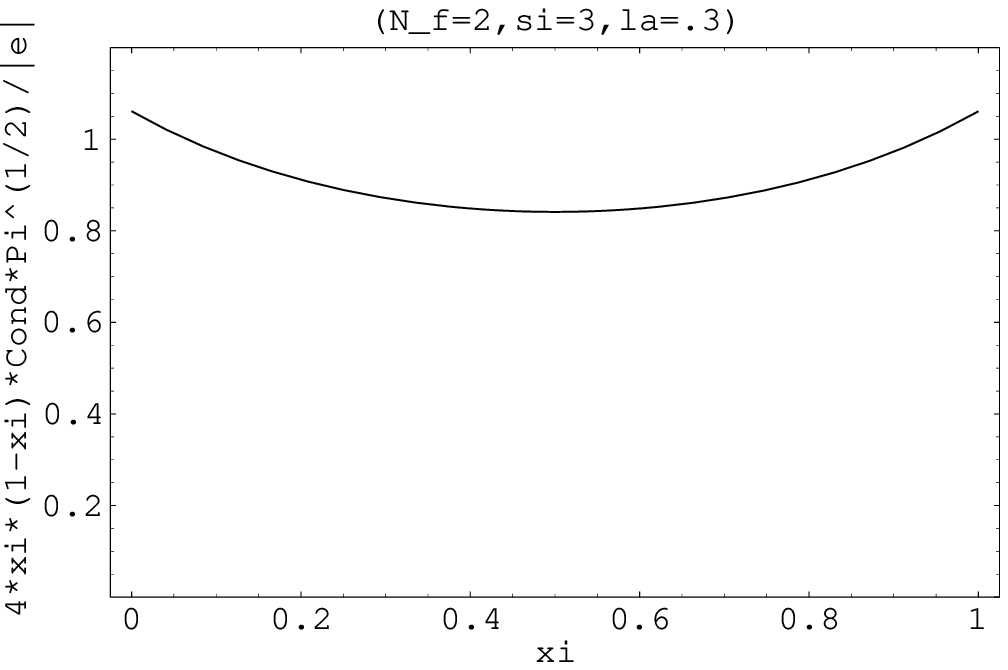,width=6.7cm,angle=90}
\end{tabular}
\caption{Spatial Dependence of $4\xi(1\!\!-\!\xi)|\<\psi\dag P_\pm\psi\>|/
\mu_1$ for $N_{\!f}=2$ ($\xi\!=\!x^1\!/L$).}
\label{fig3}
\end{figure}

We are now in a position to evaluate formulas (\ref{copc.5},\ref{copc.6})
for $N_{\!f}\!=\!1$ as well as formulas (\ref{neoc.5},\ref{neoc.6}) for
$N_{\!f}\!=\!2$.
The first hint regarding the difference between single-flavour and multi-flavour
cases might come from observing how the spatial dependence of the condensate
behaves as temperature and box-length vary.
Since the condensate diverges on the boundaries as $\xi^{-1}$ and $(1-\xi)^{-1}$
respectively 
\footnote{There is no reason to start worrying: On the boundary the field
has to satisfy a boundary condition which lets the free Green's function take
the form given in (\ref{fpro.6}). This expression stays finite as one of its
two entries --~$x$ or $y$~-- reaches the boundary but not if $x$ and $y$
reach the same point on the boundary, which simply means that there is no
consistent double-boundary solution.}, it is the quantity
$4\xi(1-\xi)|\<\psi\dag P_\pm\psi\>|/(|e|/\sqrt{\pi})$ at $\th\!=\!0$ which
is displayed in figures \ref{fig2} and \ref{fig3}.

From inspecting figures \ref{fig2} and \ref{fig3} one is lead to the
observation that for any finite temperature the condensate ends up ``creeping
into the boundaries'' (and fading away at any internal point of the box) once
the box-length is sufficiently large.

On the other hand, cooling the system seems to have the opposite effect, i.e.
to a first approximation, the shape of the spatial distribution of the
condensate seems to depend on the ratio $\ta\!=\!\sg/2\la$.
However, as $\sg$ and $\la$ both tend to be large, the spatial distribution
starts to look qualitatively different in the two cases $N_{\!f}\!=\!1$ and
$N_{\!f}\!=\!2$, respectively.
This is a pictorial hint indicating a difference between the single-flavour
and the multi-flavour versions of the model once the twofold limit
$\sg\rightarrow\infty,\la\rightarrow\infty$ is performed.


\subsection{Specialization to $N_{\!f}=1$ at Midpoints}

As one concentrates on the midpoints ($\xi=1/2$), formulae (\ref{copc.5})
and (\ref{copc.6}) take the form
\bea
{\<\psd P_{\pm}\ps\>\ov(|e|/\sqrt{\pi})}\!&\!=\!&\!
\pm{e^{\pm\th/\rch(\la/2)}\ov4\pi}
\Big(1+2\sum\limits_{n\geq1}(-1)^n\;
{1\ov\rch(n\pi\tau)}\:\exp(-n^2\pi\ta)\Big)\cdot
\nonumber
\\
&{}&\!
\exp\Big\{
\ga-2\sum\limits_{j\geq1}(-1)^jK_0(j\la)
\Big\}
\nonumber
\\
&{}&\!
\exp\Big\{4\sum\limits_{n\geq0}
{1\ov(2n\!+\!1)(e^{2(2n\!+\!1)\pi\tau}-1)}-
((2n\!+\!1)\!\rightarrow\!\sqrt{(2n\!+\!1)^2\!+\!(\la/\pi)^2\,})
\Big\}
\label{neod.1}
\\
\nonumber
\\
{\<\psd P_{\pm}\ps\>\ov(|e|/\sqrt{\pi})}\!&\!=\!&\!
\pm{e^{\pm\th/\rch(\la/2)}\ov4\pi}
\sum\limits_{m\geq0}(\!-1)^m\;
{e^{-\pi((2m+1)^2-1)/4\ta}\ov\rsh(\pi(2m+1)/2\ta)}\times
\nonumber
\\
&{}&\!
\sum\limits_{k\geq0}\rch(\mbox{${\pi(2m+1)(2k+1)\ov2\ta}$})
({\rm erf}(\mbox{${(k+1)\sqrt{\pi}\ov\sqrt{\ta}}$})
\!-\!{\rm erf}(\mbox{${(k)\sqrt{\pi}\ov\sqrt{\ta}}$}))
\cdot
\nonumber
\\
&{}&\!
\exp\Big\{\ga+{\pi(1-{\rm th}(\la/2))\ov\sg}
-2\sum\limits_{j\geq1}K_0(j\sg)
\Big\}
\cdot
\nonumber
\\
&{}&\!
\exp\Big\{
-2\sum\limits_{m\geq1}
{1\ov m(e^{m\pi/\tau}+1)}
-(m\!\rightarrow\!\sqrt{m^2\!\!+\!\!(\sg/2\pi)^2\,})
\Big\}
\label{neod.2}
\eea
respectively (for details, see appendix), from which we derive
\bea
\lim\limits_{\sg\rightarrow\infty}
{\<\psd P_{\pm}\ps\>({L\ov2})\ov(|e|/\sqrt{\pi})}\!&\!=\!&\!
\pm{e^{\pm\th/\rch(\la/2)}\ov4\pi}
\exp\Big\{
\ga-2\sum\limits_{j\geq1}(-1)^jK_0(j\la)
\Big\}
\label{neod.3}
\\
\nonumber
\\
\lim\limits_{\la\rightarrow\infty}
{\<\psd P_{\pm}\ps\>({L\ov2})\ov(|e|/\sqrt{\pi})}\!&\!=\!&\!
\pm{1\ov4\pi}
\exp\Big\{
\ga-2\sum\limits_{j\geq1}K_0(j\sg)
\Big\}
\label{neod.4}
\eea
for $N_{\!f}\!=\!1$. Thus
\beq
\lim\limits_{\la\rightarrow\infty}
\lim\limits_{\sg\rightarrow\infty}
{\<\psd P_{\pm}\ps\>({L\ov2})\ov(|e|/\sqrt{\pi})}=
\pm{1\ov4\pi}e^\ga=
\lim\limits_{\sg\rightarrow\infty}
\lim\limits_{\la\rightarrow\infty}
{\<\psd P_{\pm}\ps\>({L\ov2})\ov(|e|/\sqrt{\pi})}
\label{double.1}
\eeq
for $N_{\!f}\!=\!1$.
Note that (\ref{neod.4}) and the second equality in (\ref{double.1}) correct
for an erroneous result in \cite{DuWi} which was won by performing the limit
under the $c$-integral.


\subsection{Specialization to $N_{\!f}=2$ at Midpoints}

As one concentrates on the midpoints ($\xi=1/2$), formulae (\ref{neoc.5})
and (\ref{neoc.6}) take the form
\bea
{\<\psd P_{\pm}\ps\>\ov(|e|/\sqrt{\pi})}\!&\!=\!&\!
\pm{2^{1/4}e^{\pm\th/\rch(\la/\sqrt{2})}\ov4\sqrt{\pi\;}\sqrt{\la\;}}\cdot
\nonumber
\\
&{}&\!
\Big(1+2\sum\limits_{n\geq1}(-1)^n\;
{e^{-n^2\pi\ta/2}\ov\rch(n\pi\tau)}\cdot
{e^{-n^2\pi\ta/2}+
2\sum\limits_{k\geq1}{e^{-(n/2-k)^22\pi\ta}+e^{-(n/2+k)^22\pi\ta}\ov2}\ov
1+2\sum\limits_{k\geq1}e^{-2k^2\pi\ta}}\Big)\cdot\!
\nonumber
\\
&{}&\!
\exp\Big\{
{\ga\ov2}-\sum\limits_{j\geq1}(-1)^j K_0(j\sqrt{2}\;\la)
\Big\}
\nonumber
\\
&{}&\!
\exp\Big\{2\sum\limits_{n\geq0}
{1\ov(2n\!+\!1)(e^{2(2n\!+\!1)\pi\tau}-1)}-
((2n\!+\!1)\!\rightarrow\!\sqrt{(2n\!+\!1)^2\!+\!2(\la/\pi)^2\,})
\Big\}
\label{neod.5}
\\
\nonumber
\\
{\<\psd P_{\pm}\ps\>\ov(|e|/\sqrt{\pi})}\!&\!=\!&\!
\pm{2^{1/4}e^{\pm\th/\rch(\la/\sqrt{2})}\ov4\sqrt{\pi\;}\sqrt{\sg\;}}\;
\sum\limits_{m\geq0}(-1)^m\;
{e^{-\pi((2m+1)^2-1)/8\ta}\ov\rsh(\pi(2m+1)/2\ta)}\times
\nonumber
\\
&{}&\!
{
\begin{array}{l}
\sum\limits_{q\in Z}\!e^{-\pi q^2/2\ta}\!\sum\limits_{p\geq0}\!
\rch(\mbox{\small${\pi(p+1/2)(m+1/2)\ov\ta}$})
({\rm erf}({(p+3/2)\ov\sqrt{2\ta/\pi\;}})
\!-\!{\rm erf}({(p-1/2)\ov\sqrt{2\ta/\pi\;}}))+
\\
\sum\limits_{q\in Z}\!e^{-\pi q^2/2\ta}\!\sum\limits_{p\geq0}\!
\mbox{\small$(-1)^{p+q-m}$}\rsh(\mbox{\small${\pi(p+1/2)(m+1/2)\ov\ta}$})
({\rm erf}({(p+3/2)\ov\sqrt{2\ta/\pi\;}})
\!-\!{\rm erf}({(p-1/2)\ov\sqrt{2\ta/\pi\;}}))
\end{array}
\ov
\sum\limits_{q\in Z}e^{-\pi q^2/2\ta}
}
\nonumber
\\
&{}&\!
\exp\Big\{{\ga\ov2}
+{\pi(1\!-\!{\rm th}(\la/\sqrt{2}))\ov2\sqrt{2}\sg}
-\sum\limits_{j\geq1}K_0(j\sqrt{2\;}\sg)
\Big\}
\nonumber
\\
&{}&\!
\exp\Big\{
-\sum\limits_{m\geq1}
{1\ov m(e^{\pi m/\tau}+1)}\!-\!
(m\!\rightarrow\!\sqrt{m^2\!+\!2(\sg/2\pi)^2\,})
\Big\}
\label{neod.6}
\eea
respectively (for details, see appendix), from which we derive
\bea
\lim\limits_{\sg\rightarrow\infty}
{\<\psd P_{\pm}\ps\>({L\ov2})\ov(|e|/\sqrt{\pi})}\!&\!=\!&\!
\pm{2^{1/4}e^{\pm\th/\rch(\la/\sqrt{2})}\ov4\sqrt{\pi\;}\sqrt{\la\;}}
\exp\Big\{
{\ga\ov2}-\sum\limits_{j\geq1}(-1)^jK_0(j\sqrt{2}\la)
\Big\}
\label{neod.7}
\\
\nonumber
\\
\lim\limits_{\la\rightarrow\infty}
{\<\psd P_{\pm}\ps\>({L\ov2})\ov(|e|/\sqrt{\pi})}\!&\!=\!&\!
0
\label{neod.8}
\eea
for $N_{\!f}\!=\!2$. Thus
\beq
\lim\limits_{\la\rightarrow\infty}
\lim\limits_{\sg\rightarrow\infty}
{\<\psd P_{\pm}\ps\>({L\ov2})\ov(|e|/\sqrt{\pi})}=
0=
\lim\limits_{\sg\rightarrow\infty}
\lim\limits_{\la\rightarrow\infty}
{\<\psd P_{\pm}\ps\>({L\ov2})\ov(|e|/\sqrt{\pi})}
\label{double.2}
\eeq
for $N_{\!f}\!=\!2$.
Note that in the first case the condensate decays rather reluctantly ($\propto
1/\sqrt{\la}$) only under the outer limit, whereas in the second case it
decays exponentially fast under the inner limit already (see appendix).
This difference may be stated in equations through
\bea
\lim\limits_{\la\rightarrow\infty}
\lim\limits_{\sg\rightarrow\infty}
\sqrt{\la\;}
{\<\psd P_{\pm}\ps\>({L\ov2})\ov(|e|/\sqrt{\pi})}\!&\!=\!&\!
\pm{2^{1/4}\ov4\sqrt{\pi\;}}
\exp\Big\{
{\ga\ov2}
\Big\}
\label{double.3}
\\
\nonumber
\\
\lim\limits_{\sg\rightarrow\infty}
\lim\limits_{\la\rightarrow\infty}
\sqrt{\la\;}
{\<\psd P_{\pm}\ps\>({L\ov2})\ov(|e|/\sqrt{\pi})}\!&\!=\!&\!
0
\label{double.4}
\eea
which is peculiar for $N_{\!f}\!=\!2$.


\subsection{Numerical Evaluation at Midpoints}

We are now in a position to numerically evaluate formulas (\ref{neod.1},
\ref{neod.2}) for $N_{\!f}\!=\!1$ as well as formulas (\ref{neod.5},
\ref{neod.6}) for $N_{\!f}\!=\!2$.

The following figures display the absolute value of the condensate at
$\th\!=\!0$ in the center of the box.
The two-dimensional graphs show the condensate as a function of
$\sg\!=\!\be\cdot|e|/\pi^{1/2}$ (or a function thereof) at fixed value of
$\la\!=\!L\cdot|e|/\pi^{1/2}$. Throughout we use $\mu_1=|e|/\pi^{1/2}$.
The surface- and density-plots show the condensate versus (a function of)
$\sg$ and $\la$.

It might be worth mentioning that in each of those figures both representations
--~(\ref{neoc.1}) {\em and} (\ref{neoc.2}) in case of $N_{\!f}\!=\!1$ as well
(\ref{neoc.7}) {\em and} (\ref{neoc.8}) in case of $N_{\!f}\!=\!2$~-- were
used, the former formulas got evaluated for low temperatures ($\ta\!\gg\!1$),
the latter ones got evaluated for high temperatures ($\ta\!\ll\!1$) ---
the switching being done in the region of the crossover transition, which
essentially gives a numerical check that the representations 
(\ref{neoc.1}) and (\ref{neoc.2}) as well as the representations
(\ref{neoc.7}) and (\ref{neoc.8}) indeed might be identical.

\begin{figure}
\begin{tabular}{lr}
\epsfig{file=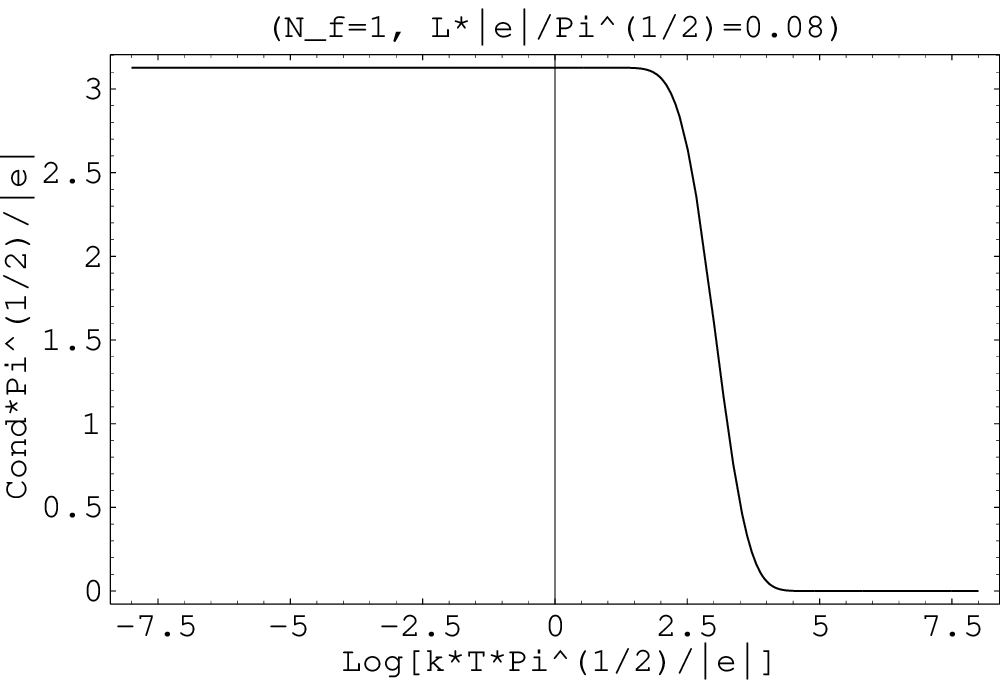,width=7cm}&
\epsfig{file=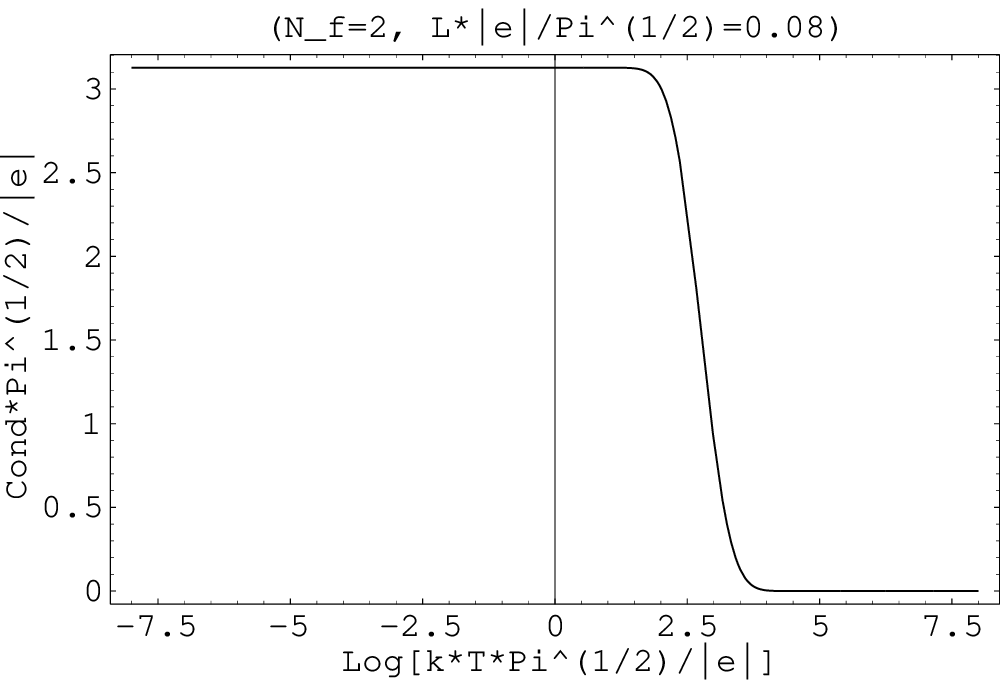,width=7cm}
\end{tabular}
\caption{$\vert\<\psi\dag P_\pm\psi\>\vert/\mu_1$ as a function of
$\log(kT/\mu_1)$ at $L=0.08/\mu_1$ for $N_{\!f}=1$ and $N_{\!f}=2$.}
\label{fig4}
\end{figure}
\begin{figure}
\begin{tabular}{lr}
\epsfig{file=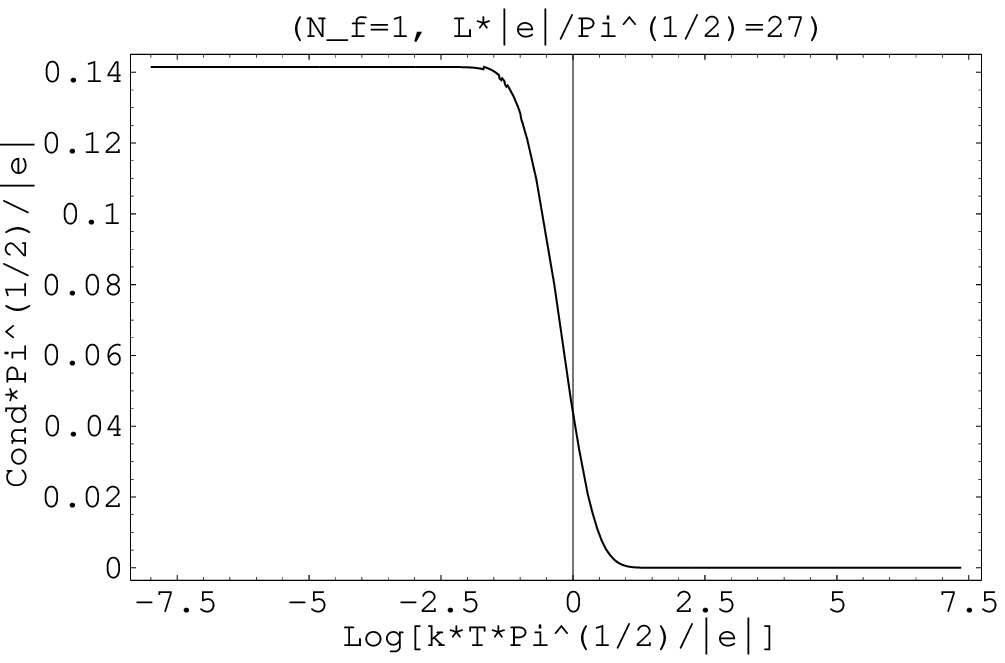,width=7cm}&
\epsfig{file=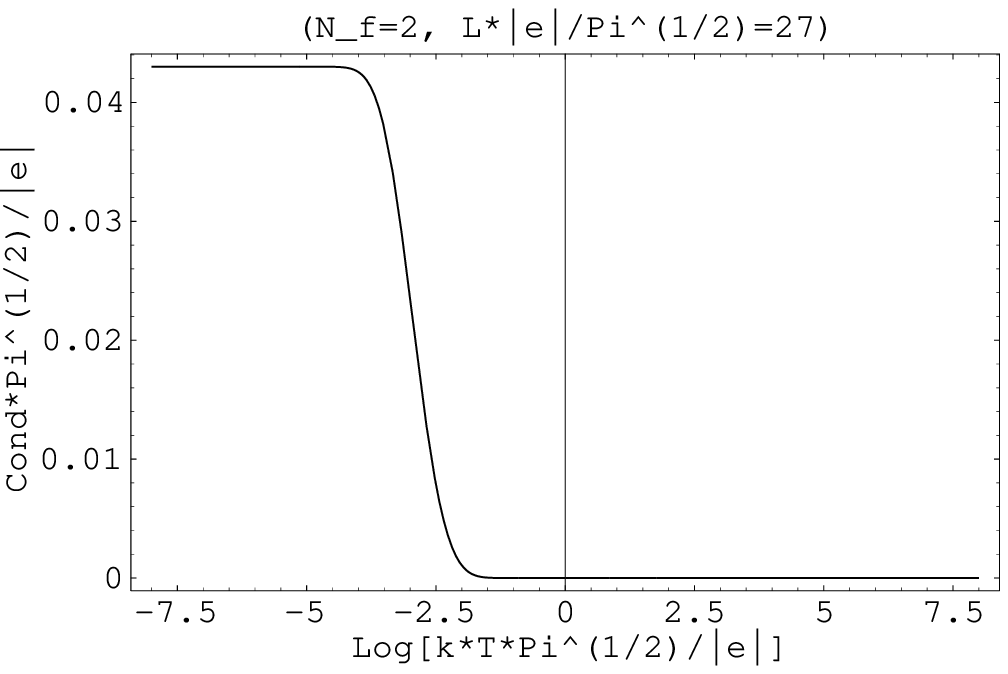,width=7cm}
\end{tabular}
\caption{$\vert\<\psi\dag P_\pm\psi\>\vert/\mu_1$ as a function of
$\log(kT/\mu_1)$ at $L=27.0/\mu_1$ for $N_{\!f}=1$ and $N_{\!f}=2$.}
\label{fig5}
\end{figure}

From figures \ref{fig4} and \ref{fig5} the system is seen to undergo a
surprisingly well-localized crossover, which --~however~-- doesn't meet the
criteria for a phase-transition of whatever kind as it is arbitrarily smooth
(i.e. $\in C^{\infty}$).

By comparing figures \ref{fig4} and \ref{fig5}, one realizes that increasing
the box-length essentially moves the ``kink'' to the left, i.e. increasing
the box-length results in a decrease of the ``critical temperature'' (the
effect being much stronger in case $N_{\!f}\!=\!2$ than for $N_{\!f}\!=\!1$)
and one starts wondering whether this kink-phenomenon survives the limit
$L\rightarrow\infty$.
We will see that the answer to this question depends in a critical way on the
number of flavours; the two cases $N_{\!f}\!=\!1$ and $N_{\!f}\!=\!2$ turn out
to be different.
The ``plateaus'' seem to be equally high for one and two flavours as long as
the box-length is small (cf. figure \ref{fig4}) but the height of the plateau
decreases unequally rapidly if the box-length is increased (cf. figure
\ref{fig5} -- compare the scales~!).

\begin{figure}
\begin{tabular}{lr}
\epsfig{file=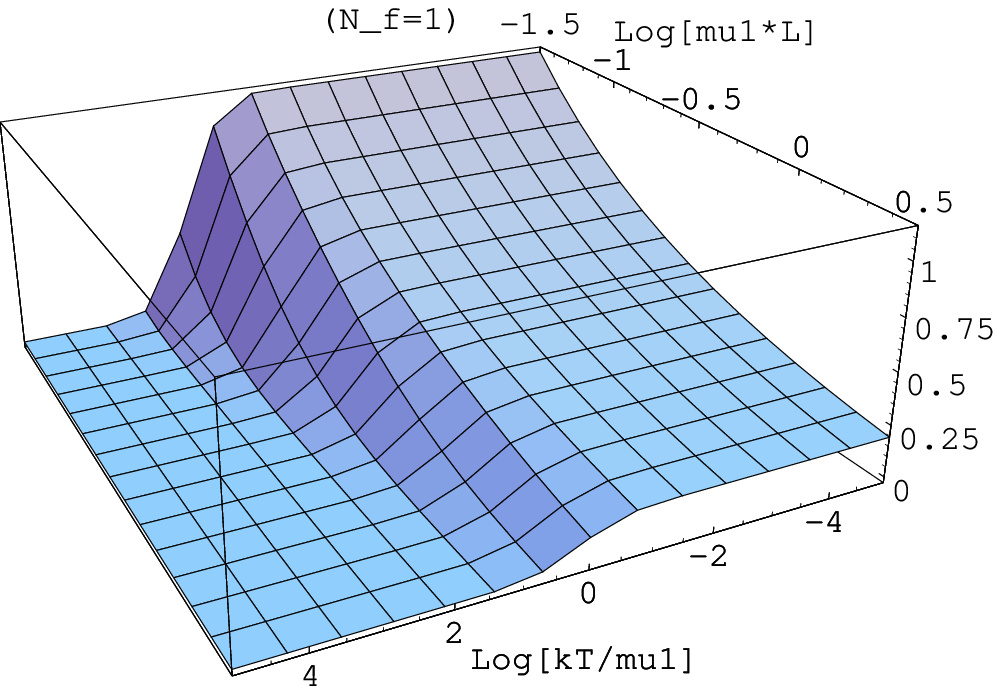,width=7cm}&
\epsfig{file=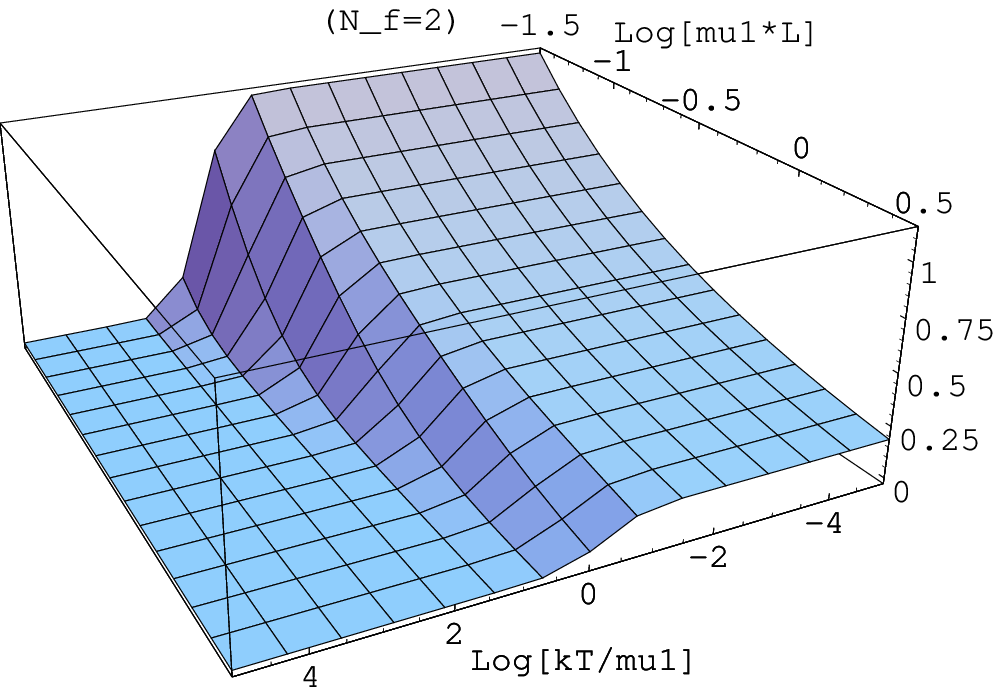,width=7cm}
\end{tabular}
\caption{$\vert\<\psi\dag P_\pm\psi\>\vert/\mu_1$ as a function of
$\log(kT/\mu_1)$ and $\log(L\mu_1)$ for $N_{\!f}=1$ and $N_{\!f}=2$.}
\label{fig6}
\end{figure}

In order to reach an intuitive understanding it is worth having a look at
figure \ref{fig6}.
Thereby one realizes that there is a considerable fraction of the
$\log(kT/\mu_1)$-$\log(L\mu_1)$-plane where the condensate is exponentially
close (but not equal) to zero.
Thus it might be helpful to introduce the concept of a {\em quasi-phase} with
{\em almost\/} restored chiral symmetry.
In order to do so one has to decide on a trigger-value which the condensate
has to exceed in order to constitute a point (in the $\log(kT/\mu_1)$-
$\log(L\mu_1)$-plane) with manifestly broken symmetry.
While this choice is --~in principle~-- arbitrary, it seems natural to agree
on half of the classical value of the condensate in the original
(one-flavour) Schwinger model as the discriminator which makes the distinction
between the two ``quasi-phases''. Numerically, it is about 0.07.

\begin{figure}
\begin{tabular}{lr}
\epsfig{file=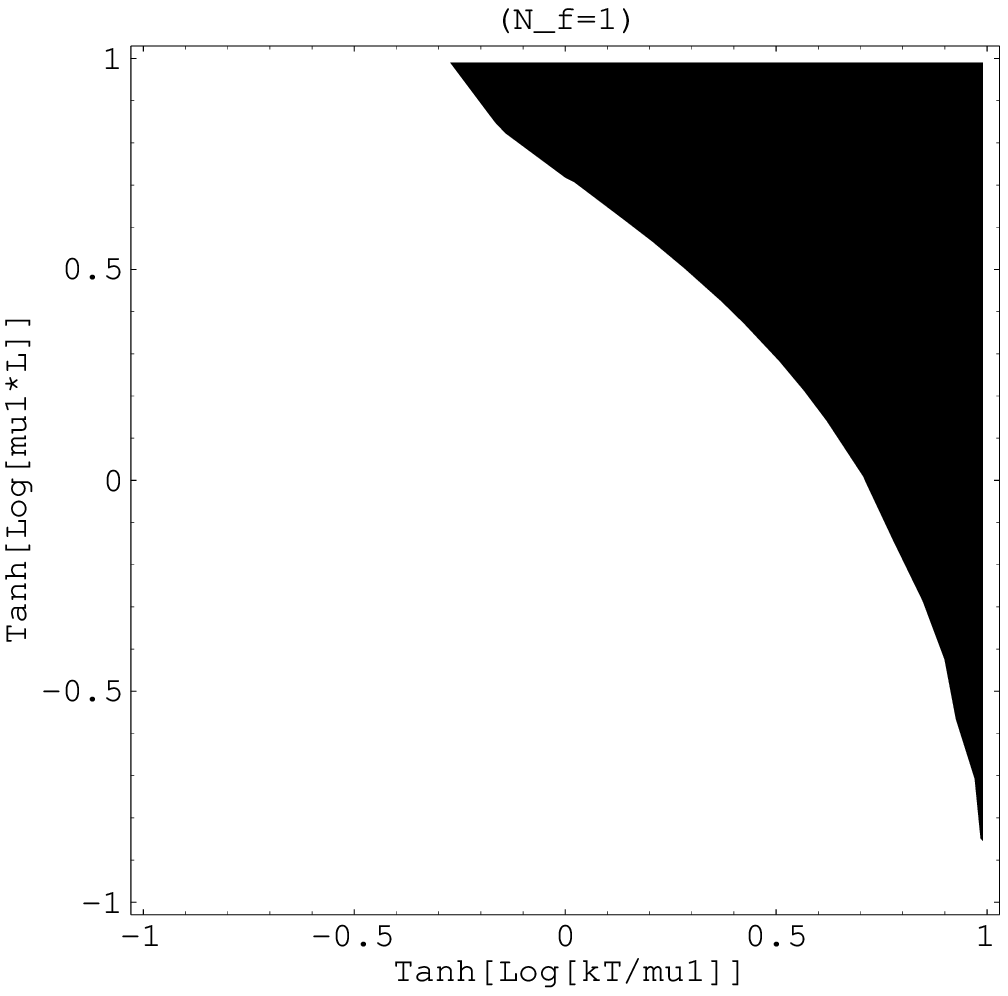,width=7cm}&
\epsfig{file=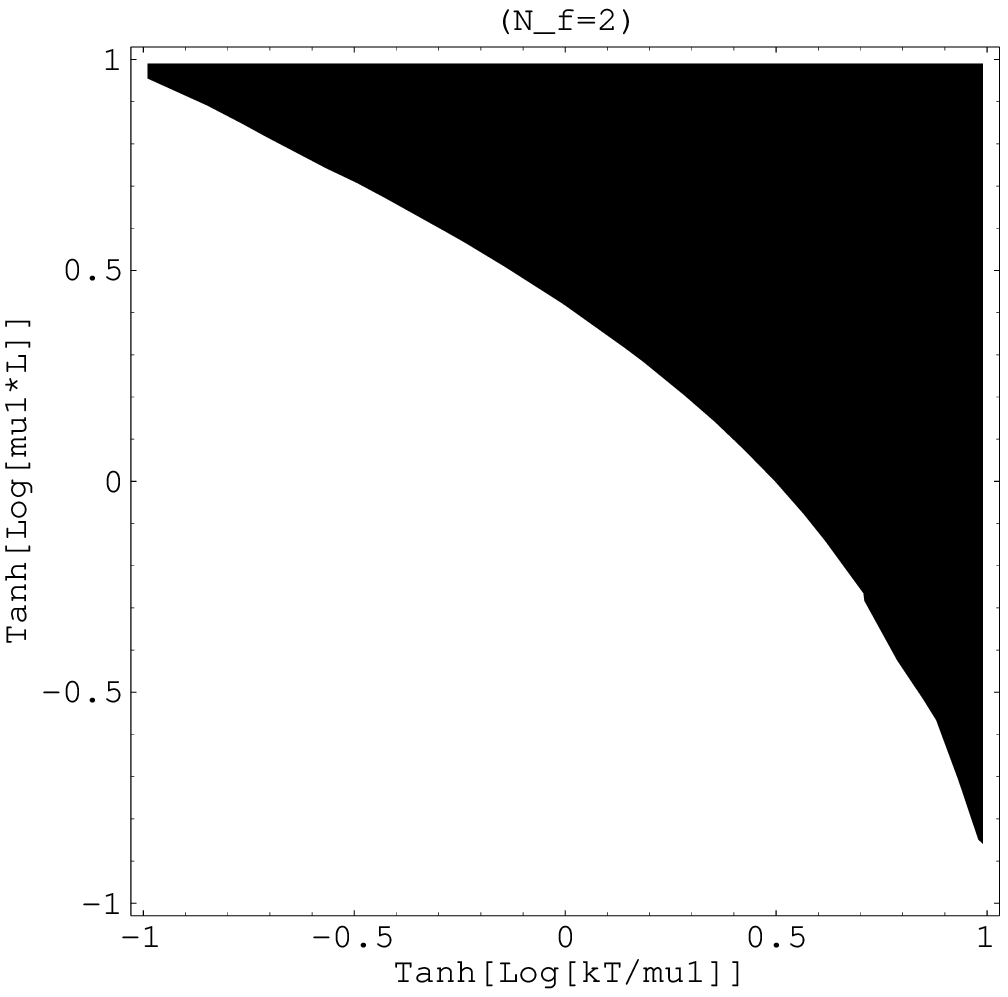,width=7cm}
\end{tabular}
\vspace{-0.2cm}
\caption{Quasi-Phasestructure as a function of $\log(kT/\mu_1)$ and
$\log(L\mu_1)$ for $N_{\!f}=1$ and $N_{\!f}=2$, respectively. Definition:
see text.}
\label{fig7}
\end{figure}

Doing so results in generating the two contour-plots in figure \ref{fig7}:
White points are those which satisfy the criterion
$\vert\<\psi\dag P_\pm\psi\>\vert \geq e^\ga\sqrt{e^2/\pi}/8\pi$ --
they constitute the (quasi-)phase with {\em manifestly broken symmetry}.
Black points are those which satisfy the criterion
$\vert\<\psi\dag P_\pm\psi\>\vert$ $\leq e^\ga\sqrt{e^2/\pi}/8\pi$ --
they constitute the quasi-phase with {\em quasi-restored symmetry}.
These two quasi-phases are separated by a ``crossover-line'' which 
essentially corresponds to the ``border'' of the zero-level plane in
figure \ref{fig6}.
In the form shown in figure \ref{fig7} the concept of quasi-phases proves
useful as it clearly shows in which areas of parameterspace the cases
$N_{\!f}\!=\!1$ and $N_{\!f}\!=\!2$ seem to be similar and in which areas
each of them shows a clearly distinct behaviour.

First we notice that the point $\be\!=\!L\!=\!0$ (lower right corner in
figure \ref{fig7}) seems to lie close to or right on the crossover-line in
either case $N_{\!f}\!=\!1$ and $N_{\!f}\!=\!2$ (note that data near the
boundaries are cut off for numerical reasons).
This observation lets us go back to formulas (\ref{copc.1}) and (\ref{copc.2})
and derive the (moderately interesting) noncommutativity-phenomenon
\bea
\lim_{ L\rightarrow 0}\lim_{\be\rightarrow 0}\
\<\ps\dag P_\pm\ps\>({L\ov2})&=&0
\qquad\qquad\ \,(\forall N_{\!f})
\label{nitl.39}
\\
\lim_{\be\rightarrow 0}\lim_{ L\rightarrow 0}\
\<\ps\dag P_\pm\ps\>({L\ov2})&=&\infty
\qquad\qquad(\forall N_{\!f})
\label{nitl.40}
\eea
which is universal for any number of flavours.

\begin{figure}
\begin{tabular}{lr}
\epsfig{file=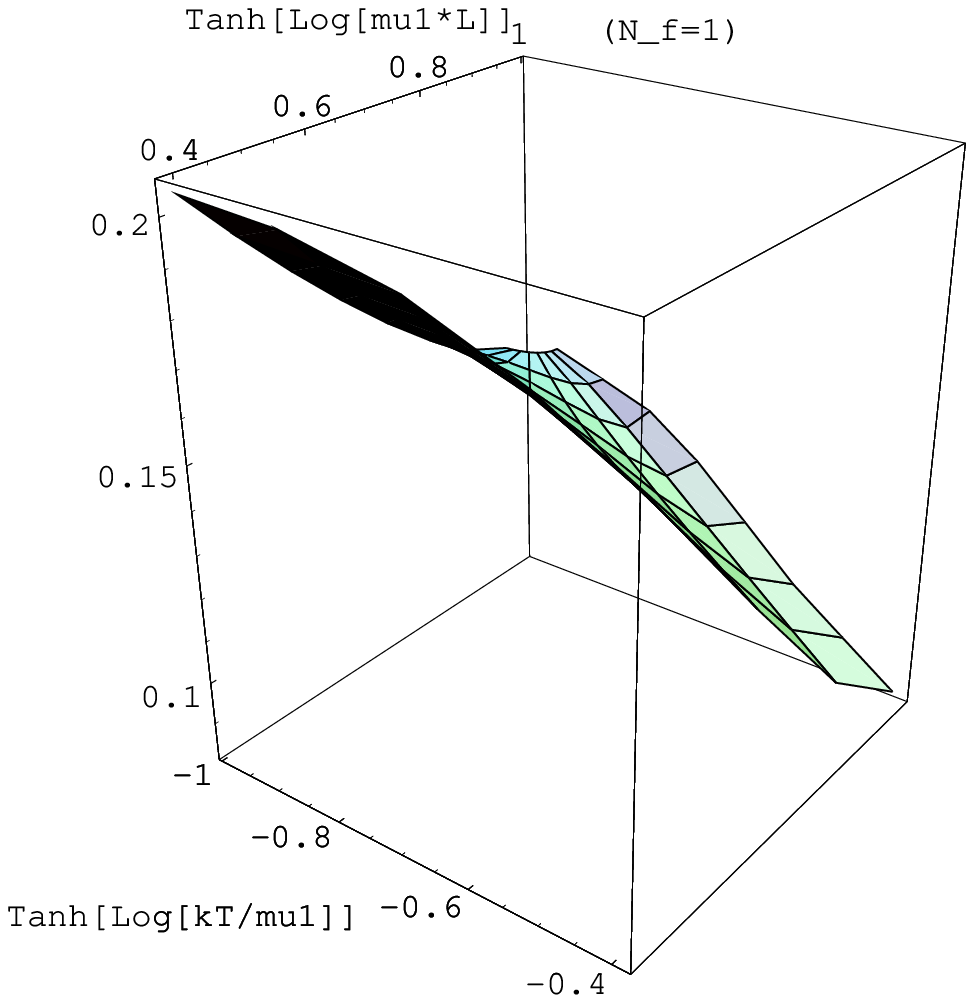,width=7cm}&
\epsfig{file=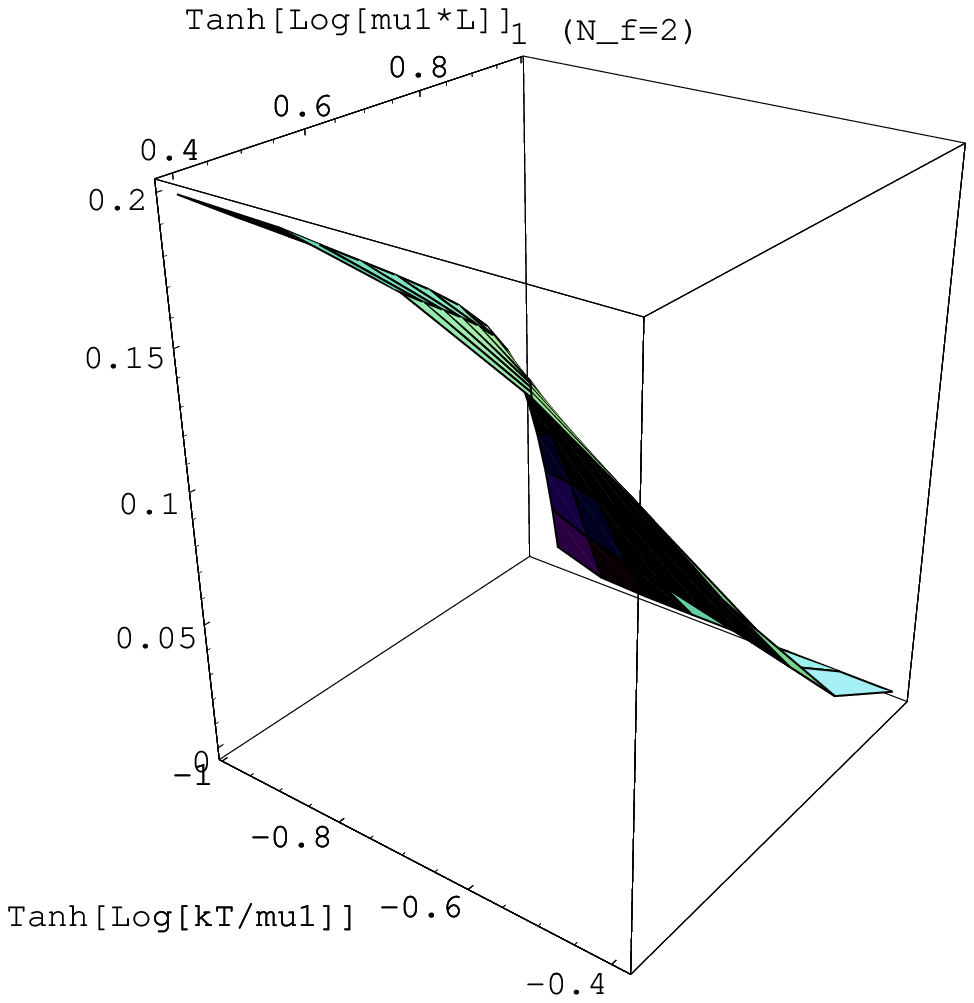,width=7cm}
\end{tabular}
\caption{Zoom-out of the area $\sg\!\gg\!1, \la\!\gg\!1$ (upper left corner in
\ref{fig7}) for $N_{\!f}\!=\!1$ and $N_{\!f}\!=\!2$.}
\label{fig8}
\end{figure}

Second we notice that the point $\be\!=\!L\!=\!\infty$ (upper left corner in
figure \ref{fig7}) definitely belongs to the manifestly broken (pseudo-)
phase for $N_{\!f}\!=\!1$ (l.h.s. of figure \ref{fig7}), but for $N_{\!f}\!=\!2$
(r.h.s. of figure \ref{fig7}) the point $\be\!=\!L\!=\!\infty$ seems either to
be part of the quasi-phase with almost restored symmetry or to lie right on the
crossover-line 
(note, again, that areas close to the boundaries in figure \ref{fig7} are cut
off for numerical reasons).
This observation lets us try to zoom into the upper left corner in these
quasi-phase-structure plots -- an attempt which results in figure \ref{fig8}.

From the l.h.s. of figure \ref{fig8} we learn that the one-flavour-system
approaches its standard-value for the order-parameter in a very
unspectacular way: The ``classical'' value for the condensate in the
single-flavour model represents a plateau which is reached smoothly
from any side.
From the r.h.s. of figure \ref{fig8} we learn that the two-flavour-system
behaves in the large-$L$-limit in a way which depends rather sensitively on
whether the temperature is exactly zero or finite: For any finite temperature
the limiting value (zero) is approached smoothly (a statement which extends
to the first derivative of the condensate), whereas at zero temperature the
condensate seems to display a ``square-root type'' behaviour (as a function of
$1/\la$).
This is exactly what was predicted by formulas (\ref{neod.7}, \ref{neod.8}):
At zero temperature the two-flavour condensate goes to zero under $\la\to\infty$
-- but only very reluctantly: $\propto 1/\sqrt{\la}$ (up to exponentially small
corrections).
On the other hand, for any fixed finite temperature, both the the condensate
and its first derivative w.r.t. $1/\la$ vanish exponentially fast under
$\la\rightarrow\infty$.

\subsection{The order of the phase-transition at $T_c\!=\!0$}

The statement that in the two-flavour model the behaviour of the chiral
condensate as a function of $\la\!=\!\mu_1L$ depends in a very sensitive way
on whether $\sg\!=\!\mu_1\be$ is finite or infinite turns out to be so crucial
in the following that one would particularly welcome some further numerical
evidence that there is, indeed, a ``square-root type'' behaviour of
the condensate for $1/\la\!\ll\!1$ at $1/\sg\!=\!0$ (as opposed to a smooth
behaviour for $\la\to\infty$ at $1/\sg\!>\!0$).

\begin{figure}
\begin{tabular}{lr}
\epsfig{file=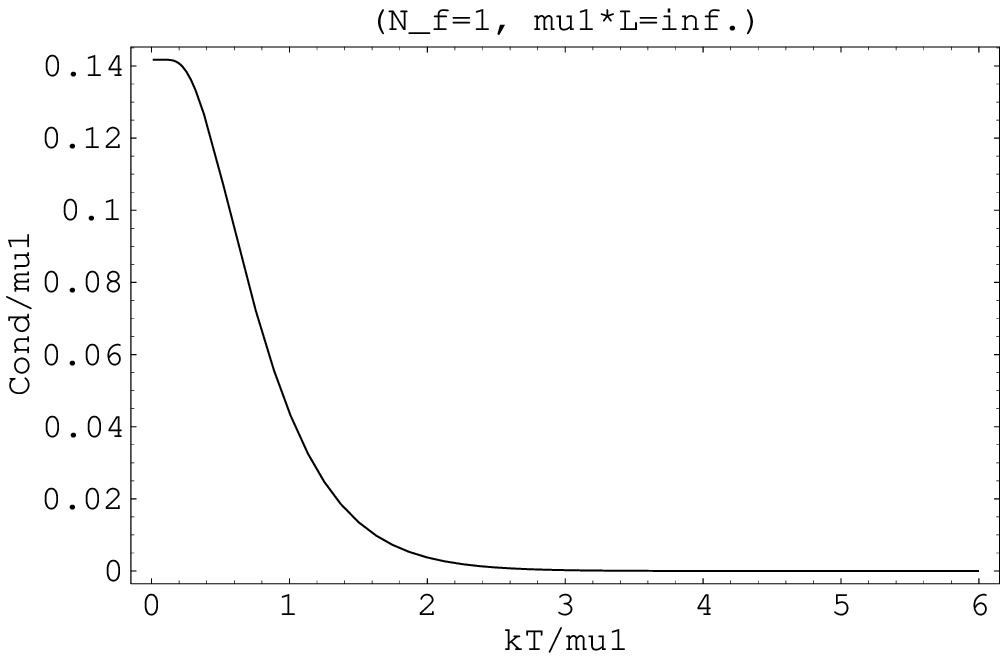,width=7cm}&
\epsfig{file=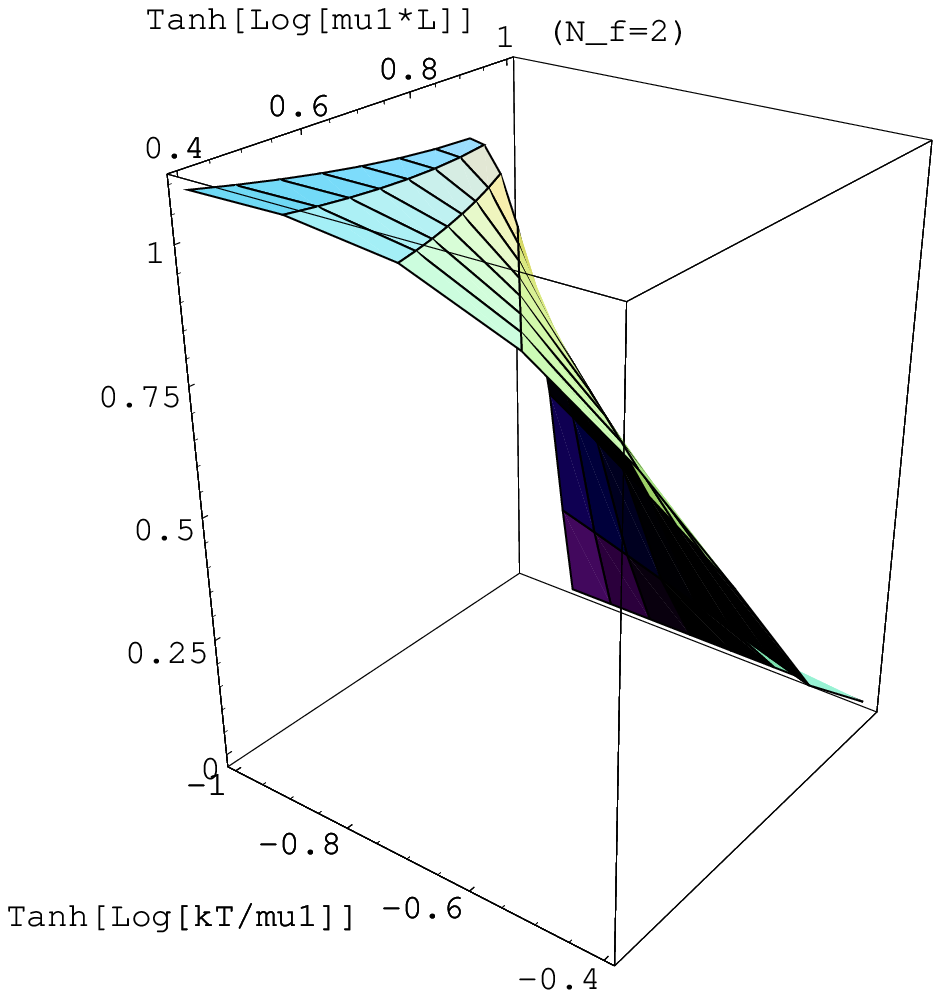,width=7cm}
\end{tabular}
\caption{Left: The order-parameter $|\<\psd P_\pm\ps\>|/\mu_1$ of the
one-flavour system as a function of the dimensionless temperature at infinite
box-length. Right: Zoom-out of the upper left corner in the r.h.s. of figure
\ref{fig7} for the two-flavour system where the alternative quantity ({\em not
an order-parameter} !)
$2^{7/4}\pi^{1/2}e^{-\ga}\cdot\la^{1/2}\vert\<\psd P_\pm\ps\>\vert/\mu_1$
is plotted rather than $\<\psd P_\pm\ps\>/\mu_1$ -- providing direct numerical
evidence for the noncommutativity phenomenon (\ref{double.3}, \ref{double.4}).}
\label{fig9}
\end{figure}

Though there is, quite generally, no numerical proof for smoothness, strong
numerical evidence can be given that the boundary associated to $T\!=\!0$
of the two-flavour condensate surface in the r.h.s. of figure \ref{fig8} does
not just look like a square-root but, indeed, asymptotically gets a square-root
and that this small-$1/\la$-behaviour is, indeed, specific for $T\!=\!0$.
To this end we simply decide to plot, for the two-flavour system, the quantity
$2^{7/4}\pi^{1/2}e^{-\ga}\cdot\sqrt{\la\;}\;\vert\<\psd P_\pm\ps\>\vert/\mu_1$,
i.e. to include a factor $\propto\sqrt{\la}$.
The result is shown in the r.h.s. of figure \ref{fig9}: the $T\!=\!0$-boundary
tends to 1 whereas the $1/L\!=\!\infty$-boundary seems to be compatible with 0.
In summary, the r.h.s. of figure \ref{fig9} provides an independent numerical
check of the analytical work (presented in the appendix) which has been done to
get from (\ref{neod.5}, \ref{neod.6}) to (\ref{neod.7}, \ref{neod.8}).

At this point a pitfall arizes:
If, for the two-flavour case, the ``alternative quantity''
$2^{7/4}\pi^{1/2}e^{-\ga}\cdot\sqrt{\la\;}\;\vert\<\psd P_\pm\ps\>\vert/\mu_1$
would be an order-parameter (actually it is {\em not} an order-parameter, as we
shall see in a moment), the very fact that it acquires a finite value (1 in our
normalization) at $\be\!=\!L\!=\!\infty$ if the zero-temperature-limit is taken
prior to the limit of infinite box-length while staying zero if the two limits
are performed in reverse order would lead us to the (false) conclusion that
at $\be\!=\!L\!=\!\infty$ the system consists of two coexisting phases with
relative weights which could be set by the way the point $\be\!=\!L\!=\!\infty$
is approached: Following the corresponding ``line of constant altitude'' in the
r.h.s. of figure \ref{fig9} the relative weight of either phase (seemingly)
could be given an arbitrary value between 0 and 1.
This means that we would reach the (incorrect) conclusion that the two-flavour
Schwinger model shows a first-order phase-transition with critical temperature
$T_c\!=\!0$.

While the statement that the two-flavour Schwinger model exhibits a
phase-transition at $T_c\!=\!0$ is indeed correct (see below), a
classification as of first order is not -- the transition is actually
{\em second order\/}.
To this end we shall first point out the fallacy in the above pseudo-reasoning
and then present the correct argument.

The first question is: what is wrong in the argument which was declared
to be a pitfall~? The answer is: nothing is wrong -- except for the fact that
the if-condition is not fulfilled.
The point is: the ``alternative quantity'' (the condensate times a factor
$\sqrt{\la}$) is {\em not} an order-parameter.
This might come as a surprise: Usually, in order to study critical phenomena
or spontaneous symmetry breaking in field theory, the prescription is that
one should choose an operator which is non-invariant under the symmetry in
question. Then, in the presence of an external symmetry-breaking field, the
vacuum-expectation-value of this operator for different temperatures and
different external fields indicates the type of the phase-transition.
In the case at hand the true order-parameter (the chiral condensate) and
the ``alternative quantity'' differ from each other by a scalar factor, i.e.
these two operators transform in the same way under a chiral transformation.
Nevertheless, the correct order-parameter and the ``alternative quantity'' do
not prove equally useful to study the chiral structure of the theory:
The ``alternative quantity'' can be seen as a specific member of a
one-dimensional class of operators where the parameter is just the exponent
of the prefactor $\la$ in front of the condensate.
Operators for which this parameter is smaller than $1/2$ tend to zero
under $\la\to\infty$ whereas operators for which this parameter is bigger
than $1/2$ diverge under $\la\to\infty$ (at zero temperature).
Demanding this parameter to equal $1/2$ produces the finite jump at
critical temperature shown in figure \ref{fig9} but there is no physical
meaning in this behaviour whatsoever:
The ``alternative quantity'' $\propto\la^{1/2}\cdot\vert\<\psd P_\pm\ps\>\vert
/\mu_1$ can not be used as an order parameter since its very definition either
requires analytical knowledge about the behaviour of the condensate (which is
the true order parameter) for large values of $\la$ at zero temperature or
so-to-say ``critical tuning'' which is completely intolerable.
Moreover, even if we had overlooked this, the (incorrect) conclusion that
the phase-transition is first order would not even be self-consistent: 
Whenever a system exhibits a true first-order transition
a sufficiently small perturbation by the
symmetry-breaking external field still leaves the transition first-order.
This, however, is definitely not true for the case at hand: The condensate
(and henceforth any operator in the one-dimensional set introduced above)
was found to show an arbitrarily smooth crossover (as a function of $\sg$)
for any nonzero value of $1/\la$ (which, in our setting, plays the role of
a small quark-mass term).

Having convinced ourselves that the condensate $\vert\<\psd P_\pm\ps\>\vert
/\mu_1$ is the unique legitimate order parameter it is straightforward to
read off the order of the phase-transition of the two-flavour system directly
from the r.h.s. of figure \ref{fig8}:
At zero temperature the condensate follows a ``square-root'' behaviour down
to its limiting value (sc. $0$) under $1/\la\to0+$ whereas for any finite
temperature it decays exponentially and henceforth such that even its
derivative tends to zero under $1/\la\to0+$.
Thus there is no finite jump and no phase-coexistence at $T_c=0$, the
susceptibility (which, in our setting, is associated with the derivative of
the condensate w.r.t. $1/\la$) diverges under $1/\la\to0+$ for $T\!=\!0$
like $1/\sqrt{(1/\la)\,}$ but tends to zero under $1/\la\to0+$ for $T\!>\!0$ and
the transition is {\em second order\/}.


\subsection{Determination of $\de$}

A system with a second order phase-transition exhibits a critical behaviour
at its vicinity which is described by a set of critical exponents.
While some of them refer to the broken phase and thus can not be defined in
the case at hand, some other describe how the system behaves at or slightly
above the transition under the influence of a symmetry-breaking external field
and may be defined even if $T_c\!=\!0$.

In order to rephrase our results in the language of the usual approach where the
infrared regulating and explicit chiral symmetry breaking device is the small
fermion mass $m$ rather than the inverse box-length $1/L$ one has to agree on
how one of these two quantities shall be translated into the other one.
It is clear that $1/L$ must be associated with a monotonic function of $m$.
Sticking to  the ``naive'' (i.e. dimensionally motivated) choice of identifying
$1/L\leftrightarrow m$, the order-parameter in the massive zero-temperature 
Schwinger model is easily derived from (\ref{neoc.10}) (for $N_{\!f}\!\geq\!2$)
\beq
\lim\limits_{T\downarrow 0}\;
\<\psd P_\pm \ps\>\propto
N_{\!f}^{1/2N_{\!f}}\;
m^{(N_{\!f}-1)/N_{\!f}}\;
\exp\Big\{{1\ov N_{\!f}}
\big(\ga-2\sum\limits_{j\geq1}(-1)^jK_0(j\sqrt{N_{\!f}}{|e|\ov m})\big)
\Big\}
\label{result.1}
\eeq
whereas at finite temperature the order parameter and its first derivative 
w.r.t. $m$ still go to zero under $m\to0+$.
From this we learn two things:

(1) The phenomenon of a second-order phase-transition with zero critical
temperature is not specific for the two-flavour model --
the (massless) Schwinger model has a phase-transition at
$T_c\!=\!0$ for {\em any number of flavours bigger than one\/}.

(2) The critical exponent $\de$ which is defined through $\<\psd P_\pm \ps\>
\propto m^{1/\de}$ (at $T\!=\!T_c$) is $N_{\!f}/(N_{\!f}\!-\!1)$, i.e.
$\de\!=\!2$ for $N_{\!f}\!=\!2$ (upon identifying $1/L\leftrightarrow m$ -- for
a critical remark see below).


\section{Discussion and Conclusion}


The present paper has been devoted to a study of the $N_{\!f}$-flavour euclidean
Schwinger model on a finite-temperature cylinder with $SU(N_{\!f})_A$ breaking
local boundary conditions at the two spatial ends.
We have investigated the value of the dimensionless condensate
$\<\ps\dag P_\pm\ps\>/\mu_1$ at midpoints --~which we used as order-parameter~--
as a function of the dimensionless inverse temperature $\sg\!=\!\mu_1\be$ and
box-length $\la\!=\!\mu_1 L$.

Our aim was to give a qualitative picture of the behaviour of the Schwinger
model when quantized as described above.
We found that on a logarithmic temperature-scale (with $L$ kept finite but
fixed) the condensate undergoes a well-localized crossover from a fairly
constant value (at low temperatures) to a value which is exponentially close
to zero at sufficiently high temperature -- both for $N_{\!f}\!=\!1$ and
$N_{\!f}\!=\!2$.
From this we concluded that it is most reasonable to distinguish a (quasi-)phase
with manifestly broken chiral symmetry from a quasi-phase where the chiral
symmetry is almost restored -- the distinction being done trough a critical
value of the condensate which was defined as half of the classical value of the
condensate in the single-flavour model at zero temperature.

The numerical illustrations of our analytical results show that there is
a qualitative difference in the behaviour of the single-flavour model
as compared to the two-flavour model if the combined limit of large box-length
and small temperature is considered.
For $N_{\!f}\!=\!1$ the condensate takes a nonzero ($T$-dependent) value after
$L\rightarrow\infty$.
For $N_{\!f}\!=\!2$ the condensate ends up being exactly zero under
$L\rightarrow\infty$, but the way it approaches this limiting value is rather
different for fixed $T\!=\!0$ versus fixed $T\!>\!0$: extremely reluctantly
(i.e. $\propto 1/\sqrt{L}$) in the first case versus exponentially fast in the
second case.
This means that the susceptibility (defined as the first derivative of the
condensate w.r.t. $1/L$) diverges in the two-flavour model like $\sqrt{L}$
under $L\to\infty$ at zero temperature but tends to zero under $L\to\infty$
at any fixed finite temperature.
This means that the two-flavour system exhibits a second-order phase-transition
with critical temperature $T_c\!=\!0$ -- a result which was shown to extend to
the cases $N_{\!f}\!\geq\!3$.
For $N_{\!f}\!=\!1$, on the other hand, the order-parameter stays nonzero after
$L\rightarrow\infty$ for any $T$, and the fact that it is just exponentially
close but not equal to zero for $T\gg|e|$ illustrates the statement by Dolan and
Jackiw that the one-flavour system never restores the anomalously broken
$U(1)_A$-symmetry at finite temperature \cite{DoJa}.

It is worth emphasizing that the results summarized so far provide direct
pictorial evidence for the claim by Smilga and Verbaarschot \cite{SmVe} which
was based on an interesting indirect argument:
These two authors used the result for the scalar susceptibility in the Schwinger
model with degenerate massive flavours determined via bosonization rules and
completed the list of critical exponents at small but nonzero temperature.
Their result was that this list can be understood most easily as to consist of
entries which satisfy the scaling relations for a system slightly above a
second-order phase-transition with $T_c\!=\!0$.

There is, however, one important point of numerical disagreement:
Based on our analytical result plus the identification rule $m\leftrightarrow
1/L$ we have found the critical exponent $\de\!=\!2$ for the two-flavour
system -- which is at variance with the bosonization-rule based result
$\de\!=\!3$ (which also agrees with the mean-field value) found in the
literature \cite{SmVe}.

As far as our part is concerned all we can say is that we have tried most
diligently to make sure that our result is not due to an error in the
analytical computations: 
Our finding for the numerical value of $\de$ in the two-flavour case stems
from (\ref{neod.7}) which was derived from (\ref{neod.5}). Through an 
explicit plot (figure \ref{fig9}) we tried to convince ourselves numerically
that all the subleading factors in (\ref{neod.5}) do indeed get marginal
in the zero-temperature limit and the overall $\la$-dependence tends to
$1/\sqrt{\la}$ (for $N_{\!f}\!=\!2$).
Moreover numerical evaluations of (\ref{neod.5}) and (\ref{neod.6}) (after
having cut down the infinite sums to an appropriate finite number of terms)
were found to agree to more than a dozen decimal places for a large variety
of $\sg$- and $\la$-values.

As far as the result for $\de$ in the literature is concerned all we can say is
that the results for the massive Schwinger model we are aware of are based
either on perturbation-theory in the fermion-mass \cite{Adam} or on
bosonization-rules (see \cite{BosMassSchwi} and references therein).
While the original derivation of some of these rules was again within the
mass-perturbation approach \cite{Bosonization}, the rules themselves seem
not to be tied to perturbation-theory:
Hetrick, Hosotani and Iso found in the bosonized massive Schwinger model with
2 or 3 degenerate flavours a noncommutativity-phenomenon between $m\to 0$ and
$L\to\infty$ (where $L$ plays in their scheme the role of a finite inverse
temperature $\be$) \cite{HeHoIs}, which would be rather surprising if the
bosonization-rules would not go beyond mass-perturbation theory.
Nevertheless, for the bosonization rules to be applicable the condition
$m\ll|e|$ has to hold true, whereas our results stem from an approximation-free
analytical computation and are supposed to be exact for any inverse box-length
and temperature.

At this point it should be stressed that the identification $1/\la
\leftrightarrow m$ (which was necessary in order to extract, from our results,
a value for $\de$ in the usual approach where the symmetry breaking field
is the quark-mass rather then the inverse box-length) is not canonical;
a finite-volume effect might, in principle, spoil the validity of our
simple dimesionally motivated identification rule.
Though this possibility seems rather unlikely, we feel that in the light of the
known peculiarities of the two-dimensional world (the classical one being
described in \cite{Coleman}) it can't be ruled out a priori. 
Needless to say that in the present situation both further analytical results
in the massive multi-flavour Schwinger model and a determination of its
critical indices from the lattice would be highly desirable.
However, the fact that the critical temperature is zero provides a sort of a
challenge for the lattice approach as it requires sophisticated reweighting
techniques.

In summary we have presented a study of the chiral condensate in the one-
and two-flavour finite-temperature Schwinger model in a quantization scheme
where the usual quark-mass term is replaced by bag-inspired chiral symmetry
breaking boundary conditions.
Unlike results won from bosonization-rules or within mass-perturbation
our formulas represent (hopefully) exact analytical findings gained through a
straightforward evaluation of the path-integral.
We have introduced the concept of  {\em quasi-phases\/} in order to distinguish
regions in the parameter-space where the symmetry under investigation is
{\em manifestly broken\/} from those where it is {\em almost restored\/} and
we have provided a direct pictorial verification of the claim by Smilga and
Verbaarschot that the two-flavour Schwinger model undergoes a phase-transition
at $T_c\!=\!0$ and that the transition is of second order \cite{SmVe}.
At the time being we are unable to resolve the discrepancy between our
value for the critical exponent $\de\!=\!N_{\!f}/(N_{\!f}\!-\!1)$ and the
bosonization-rule based result $\de\!=\!(N_{\!f}\!+\!1)/(N_{\!f}\!-\!1)$.


\subsection*{Acknowledgments}

It is a pleasure to acknowledge the hospitality received at the Institute
for theoretical Physics of the University of Z\"{u}rich where parts of this work
were done.
In addition, I would like to thank A.Wipf for a previous collaboration from
which our investigation has taken benefit and an anonymous referee for
pointing out an error in an earlier version of the manuscript.
\newline
This work has been supported by the Swiss National Science Foundation (SNF).


\section*{Appendix}

Here we shall give the details for the derivation of formulas (\ref{neod.1},
\ref{neod.2}, \ref{neod.5}, \ref{neod.6}).


As one concentrates on the midpoints ($\xi=1/2$), the r.h.s. of (\ref{copc.5})
takes the form
\bea
\mbox{(\ref{copc.5})}
\!&\!=\!&\!
\pm{e^{\pm\th/\rch(\la/2)}\ov4\la}
\sum\limits_{n\in Z}(-1)^n\;
{1\ov\rch(n\pi\tau)}\:\exp(-n^2\pi\ta)\cdot
\nonumber
\\
&{}&\!
\exp\Big\{2\sum\limits_{n\geq0}
{{\rm cth}((2n\!+\!1)\pi\tau)\ov(2n\!+\!1)}-
((2n\!+\!1)\!\rightarrow\!\sqrt{(2n\!+\!1)^2\!+\!(\la/\pi)^2\,})
\Big\}
\nonumber
\\
\nonumber
\\
&\!=\!&\!
\pm{e^{\pm\th/\rch(\la/2)}\ov4\la}
\Big(1+2\sum\limits_{n\geq1}(-1)^n\;
{1\ov\rch(n\pi\tau)}\:\exp(-n^2\pi\ta)\Big)\cdot
\nonumber
\\
&{}&\!
\exp\Big\{2\sum\limits_{n\geq0}
{1\ov(2n\!+\!1)}-
{1\ov\sqrt{(2n\!+\!1)^2\!+\!(\la/\pi)^2\,}}
\Big\}
\nonumber
\\
&{}&\!
\exp\Big\{4\sum\limits_{n\geq0}\sum\limits_{m\geq1}
{e^{-2m(2n\!+\!1)\pi\tau}\ov(2n\!+\!1)}-
((2n\!+\!1)\!\rightarrow\!\sqrt{(2n\!+\!1)^2\!+\!(\la/\pi)^2\,})
\Big\}
\nonumber
\\
\nonumber
\\
&\!=\!&\!
\pm{e^{\pm\th/\rch(\la/2)}\ov4\la}
\Big(1+2\sum\limits_{n\geq1}(-1)^n\;
{1\ov\rch(n\pi\tau)}\:\exp(-n^2\pi\ta)\Big)\cdot
\nonumber
\\
&{}&\!
\exp\Big\{
2\Big({\ga\ov2}+\ha\log({\la\ov\pi})-\sum\limits_{j\geq1}(-1)^jK_0(j\la)\Big)
\Big\}
\nonumber
\\
&{}&\!
\exp\Big\{4\sum\limits_{n\geq0}
{1\ov(2n\!+\!1)(e^{2(2n\!+\!1)\pi\tau}-1)}-
((2n\!+\!1)\!\rightarrow\!\sqrt{(2n\!+\!1)^2\!+\!(\la/\pi)^2\,})
\Big\}
\label{neoc.1}
\eea
which is the result (\ref{neod.1}) quoted in subsection 4.5 and where we used
\cite{GrRy}
\beq
\sum\limits_{j\geq0}{1\ov 2j\!+\!1}-{1\ov\sqrt{(2j\!+\!1)^2+(x/\pi)^2}}=
{\ga\ov2}+\ha\log({x\ov\pi})-\sum\limits_{j\geq1}(-1)^jK_0(jx)
\label{math.1}
\eeq
with $\ga=0.57\ldots$ the Euler gamma and $K_0$ the modified Bessel function.


As one concentrates on the midpoints ($\xi=1/2$), the r.h.s. of (\ref{copc.6})
takes the form
\bea
\mbox{(\ref{copc.6})}
\!&\!=\!&\!
\pm{e^{\pm\th/\rch(\la/2)}\ov2\sg}
\sum\limits_{m\in Z}(\!-1)^m\;
{1\ov\rsh(\pi(2m+1)/2\ta)}\times
\nonumber
\\
&{}&\!
{\sum\limits_{k\in Z}
e^{\pi(m+1/2)(2k+m+1/2)/\ta}
({\rm erf}({(k+m+1)\sqrt{\pi}\ov\sqrt{\ta}})
\!-\!{\rm erf}({(k+m)\sqrt{\pi}\ov\sqrt{\ta}}))
\ov2
}\cdot
\nonumber
\\
&{}&\!
\exp\Big\{\pi\Big({1\ov4\ta}\!-\!{{\rm th}(\la/2)\ov\sg}\Big)+
\sum\limits_{m\geq1}
{{\rm th}(m\pi/2\tau)\ov m}-
(m\!\rightarrow\!\sqrt{m^2\!\!+\!\!(\sg/2\pi)^2\,})\Big\}\qquad
\nonumber
\\
\nonumber
\\
&\!=\!&\!
\pm{e^{\pm\th/\rch(\la/2)}\ov2\sg}
\sum\limits_{m\in Z}(\!-1)^m\;
{1\ov\rsh(\pi(2m+1)/2\ta)}\times
\nonumber
\\
&{}&\!
{\sum\limits_{k\in Z}
e^{\pi(m+1/2)(2k+1-m-1/2)/\ta}
({\rm erf}({(1+(2k+1))\sqrt{\pi}\ov2\sqrt{\ta}})
+{\rm erf}({(1-(2k+1))\sqrt{\pi}\ov2\sqrt{\ta}}))
\ov2
}\cdot
\nonumber
\\
&{}&\!
\exp\Big\{\pi\Big({1\ov4\ta}\!-\!{{\rm th}(\la/2)\ov\sg}\Big)+
\sum\limits_{m\geq1}
{1\ov m}-{1\ov\sqrt{m^2\!\!+\!\!(\sg/2\pi)^2\,}}\Big\}
\cdot
\nonumber
\\
&{}&\!
\exp\Big\{
2\sum\limits_{m\geq1}
\sum\limits_{n\geq1}
(-1)^n{e^{-2n(m\pi/2\tau)}\ov m}
-(m\!\rightarrow\!\sqrt{m^2\!\!+\!\!(\sg/2\pi)^2\,})
\Big\}
\nonumber
\\
\nonumber
\\
&\!=\!&\!
\pm{e^{\pm\th/\rch(\la/2)}\ov\sg}
\sum\limits_{m\geq0}(\!-1)^m\;
{e^{-\pi((2m+1)^2-1)/4\ta}\ov\rsh(\pi(2m+1)/2\ta)}\times
\nonumber
\\
&{}&\!
\sum\limits_{k\geq0}\rch(\mbox{${\pi(2m+1)(2k+1)\ov2\ta}$})
({\rm erf}(\mbox{${((2k+1)+1)\sqrt{\pi}\ov2\sqrt{\ta}}$})
\!-\!{\rm erf}(\mbox{${((2k+1)-1)\sqrt{\pi}\ov2\sqrt{\ta}}$}))
\cdot
\nonumber
\\
&{}&\!
\exp\Big\{\ga+\log({\sg\ov4\pi})+{\pi(1-{\rm th}(\la/2))\ov\sg}
-2\sum\limits_{j\geq1}K_0(j\sg)
\Big\}
\cdot
\nonumber
\\
&{}&\!
\exp\Big\{
-2\sum\limits_{m\geq1}
{1\ov m(e^{m\pi/\tau}+1)}
-(m\!\rightarrow\!\sqrt{m^2\!\!+\!\!(\sg/2\pi)^2\,})
\Big\}
\label{neoc.2}
\eea
which is the result (\ref{neod.2}) quoted in subsection 4.5 and where we used
\cite{GrRy}
\beq
\sum\limits_{j\geq1}{1\ov j}-{1\ov\sqrt{j^2+(x/\pi)^2}}=
\ga+{\pi\ov 2x}+\log({x\ov2\pi})-2\sum\limits_{j\geq1}K_0(2jx)
\label{math.2}
\eeq
with $\ga=0.57\ldots$ the Euler gamma and $K_0$ the modified Bessel function.


As one concentrates on the midpoints ($\xi=1/2$), the r.h.s. of (\ref{neoc.5})
takes the form
\bea
\mbox{(\ref{neoc.5})}
\!&\!=\!&\!
\pm{e^{\pm\th/\rch(\la/\sqrt{2})}\ov4\la}
\sum\limits_{n\in Z}(-1)^n\;{1\ov\rch(n\pi\tau)}\cdot
{\sum\limits_{k\in Z}e^{-k^2\pi\ta-(n-k)^2\pi\ta}\ov
\sum\limits_{k\in Z}e^{-2k^2\pi\ta}}\cdot
\nonumber
\\
&{}&\!
\exp\Big\{\sum\limits_{n\geq0}
{{\rm cth}((2n\!+\!1)\pi\tau)\ov(2n\!+\!1)}-
((2n\!+\!1)\rightarrow\sqrt{(2n\!+\!1)^2+2(\la/\pi)^2\,})\Big\}
\nonumber
\\
\nonumber
\\
&\!=\!&\!
\pm{e^{\pm\th/\rch(\la/\sqrt{2})}\ov4\la}
\sum\limits_{n\in Z}(-1)^n\;{1\ov\rch(n\pi\tau)}\cdot
{e^{-n^2\pi\ta/2}
\sum\limits_{k\in Z}{e^{-(n/2-k)^22\pi\ta}+e^{-(n/2+k)^22\pi\ta}\ov2}
\ov
\sum\limits_{k\in Z}e^{-2k^2\pi\ta}}\cdot
\nonumber
\\
&{}&\!
\exp\Big\{\sum\limits_{n\geq0}
{1\ov(2n\!+\!1)}-
{1\ov\sqrt{(2n\!+\!1)^2\!+\!2(\la/\pi)^2\,}}
\Big\}
\nonumber
\\
&{}&\!
\exp\Big\{2\sum\limits_{n\geq0}\sum\limits_{m\geq1}
{e^{-2m(2n\!+\!1)\pi\tau}\ov(2n\!+\!1)}-
((2n\!+\!1)\!\rightarrow\!\sqrt{(2n\!+\!1)^2\!+\!2(\la/\pi)^2\,})
\Big\}
\nonumber
\\
\nonumber
\\
&\!=\!&\!
\pm{e^{\pm\th/\rch(\la/\sqrt{2})}\ov4\la}
\Big(1+2\sum\limits_{n\geq1}(-1)^n\;
{e^{-n^2\pi\ta/2}\ov\rch(n\pi\tau)}\cdot
{e^{-n^2\pi\ta/2}\!+\!
2\!\sum\limits_{k\geq1}{e^{-(n/2-k)^22\pi\ta}+e^{-(n/2+k)^22\pi\ta}\ov2}
\ov
1+2\sum\limits_{k\geq1}e^{-2k^2\pi\ta}}\Big)\cdot\!
\nonumber
\\
&{}&\!
\exp\Big\{
{\ga\ov2}+\ha\log({\sqrt{2\;}\la\ov\pi})-
\sum\limits_{j\geq1}(-1)^j K_0(j\sqrt{2}\;\la)
\Big\}
\nonumber
\\
&{}&\!
\exp\Big\{2\sum\limits_{n\geq0}
{1\ov(2n\!+\!1)(e^{2(2n\!+\!1)\pi\tau}-1)}-
((2n\!+\!1)\!\rightarrow\!\sqrt{(2n\!+\!1)^2\!+\!2(\la/\pi)^2\,})
\Big\}
\label{neoc.7}
\eea
which is the result (\ref{neod.5}) quoted in subsection 4.5. and where we
used (\ref{math.1}).


As one concentrates on the midpoints ($\xi=1/2$), the r.h.s. of (\ref{neoc.6})
takes the form
\bea
\mbox{(\ref{neoc.6})}
\!&\!=\!&\!
\pm{e^{\pm\th/\rch(\la/\sqrt{2})}\ov2\sg}\sum\limits_{m\in Z}(-1)^m\;
{1\ov\rsh(\pi(2m+1)/2\ta)}\times
\nonumber
\\
&{}&\!
{\sum\limits_{p\in Z}\sum\limits_{q\in Z}
{1+(-1)^{p+q}\ov2}\;
e^{\pi((p+m+1/2)^2-p^2-q^2)/2\ta}
({\rm erf}({(p+m+3/2)\ov\sqrt{2\ta/\pi\;}})
-{\rm erf}({(p+m-1/2)\ov\sqrt{2\ta/\pi\;}}))
\ov
2\sum\limits_{q\in Z}
e^{-\pi q^2/2\ta}\
}\cdot\!
\nonumber
\\
&{}&\!
\exp\Big\{{\pi\ov2}\Big({1\ov4\ta}\!-\!
{{\rm th}(\la/\sqrt{2})\ov\sqrt{2}\sg}\Big)
+{1\ov2}\!\sum\limits_{m\geq1}\!
{{\rm th}(m\pi/2\tau)\ov m}\!-\!
(m\!\rightarrow\!\sqrt{m^2\!+\!2(\sg/2\pi)^2\,})\Big\}
\nonumber
\\
\nonumber
\\
\!&\!=\!&\!
\pm{e^{\pm\th/\rch(\la/\sqrt{2})}\ov2\sg}\sum\limits_{m\in Z}(-1)^m\;
{e^{-\pi(m+1/2)^2/2\ta}\ov\rsh(\pi(2m+1)/2\ta)}\times
\nonumber
\\
&{}&\!
{
\sum\limits_{q\in Z}\!e^{-\pi q^2/2\ta}\!\!\sum\limits_{p\in Z}\!
\mbox{\small${1+(-1)^{p+q-m}\ov2}$}
e^{\pi(p+1/2)(m+1/2)/\ta}
({\rm erf}({1+(p+1/2)\ov\sqrt{2\ta/\pi\;}})
\!+\!{\rm erf}({1-(p+1/2)\ov\sqrt{2\ta/\pi\;}}))
\ov
2\sum\limits_{q\in Z}e^{-\pi q^2/2\ta}
}\cdot
\nonumber
\\
&{}&\!
\exp\Big\{{\pi\ov2}\Big({1\ov4\ta}\!-\!
{{\rm th}(\la/\sqrt{2})\ov\sqrt{2}\sg}\Big)
+\ha\sum\limits_{m\geq1}{1\ov m}-{1\ov\sqrt{m^2+2(\sg/2\pi)^2\;}}
\Big\}
\nonumber
\\
&{}&\!
\exp\Big\{
\sum\limits_{m\geq1}\sum\limits_{n\geq1}(-1)^n
{e^{-2n(m\pi/2\tau)}\ov m}\!-\!
(m\!\rightarrow\!\sqrt{m^2\!+\!2(\sg/2\pi)^2\,})
\Big\}
\nonumber
\\
\nonumber
\\
\!&\!=\!&\!
\pm{e^{\pm\th/\rch(\la/\sqrt{2})}\ov2\sg}\sum\limits_{m\in Z}(-1)^m\;
{e^{-\pi((2m+1)^2-1)/8\ta}\ov\rsh(\pi(2m+1)/2\ta)}\times
\nonumber
\\
&{}&\!
{
\begin{array}{l}
\sum\limits_{q\in Z}\!e^{-\pi q^2/2\ta}\!\sum\limits_{p\geq0}\!
\rch(\mbox{\small${\pi(p+1/2)(m+1/2)\ov\ta}$})
({\rm erf}({(p+1/2)+1\ov\sqrt{2\ta/\pi\;}})
\!-\!{\rm erf}({(p+1/2)-1\ov\sqrt{2\ta/\pi\;}}))+
\\
\sum\limits_{q\in Z}\!e^{-\pi q^2/2\ta}\!\sum\limits_{p\geq0}\!
\mbox{\small$(-1)^{p+q-m}$}\rsh(\mbox{\small${\pi(p+1/2)(m+1/2)\ov\ta}$})
({\rm erf}({(p+1/2)+1\ov\sqrt{2\ta/\pi\;}})
\!-\!{\rm erf}({(p+1/2)-1\ov\sqrt{2\ta/\pi\;}}))
\end{array}
\ov
2\sum\limits_{q\in Z}e^{-\pi q^2/2\ta}
}
\nonumber
\\
&{}&\!
\exp\Big\{-{\pi{\rm th}(\la/\sqrt{2})\ov2\sqrt{2}\sg}
+\ha\Big(\ga+{\pi\ov\sqrt{2\;}\sg}+\log({\sqrt{2\;}\sg\ov4\pi})
-2\sum\limits_{j\geq1}K_0(j\sqrt{2\;}\sg)\Big)
\Big\}
\nonumber
\\
&{}&\!
\exp\Big\{
-\sum\limits_{m\geq1}
{1\ov m(e^{\pi m/\tau}+1)}\!-\!
(m\!\rightarrow\!\sqrt{m^2\!+\!2(\sg/2\pi)^2\,})
\Big\}
\label{neoc.8}
\eea
which is the result (\ref{neod.6}) quoted in subsection 4.5. and where we used
(\ref{math.2}).

For later use we finally mention that
\bea
&{}&
\sum\limits_{m\geq0}(-1)^m\;
{e^{-\pi((2m+1)^2-1)/8\ta}\ov\rsh(\pi(2m+1)/2\ta)}\times
\nonumber
\\
&{}&\!
{
\begin{array}{l}
\sum\limits_{q\in Z}\!e^{-\pi q^2/2\ta}\!\sum\limits_{p\geq0}\!
\rch(\mbox{\small${\pi(p+1/2)(m+1/2)\ov\ta}$})
({\rm erf}({(p+1/2)+1\ov\sqrt{2\ta/\pi\;}})
\!-\!{\rm erf}({(p+1/2)-1\ov\sqrt{2\ta/\pi\;}}))+
\\
\sum\limits_{q\in Z}\!e^{-\pi q^2/2\ta}\!\sum\limits_{p\geq0}\!
\mbox{\small$(-1)^{p+q-m}$}\rsh(\mbox{\small${\pi(p+1/2)(m+1/2)\ov\ta}$})
({\rm erf}({(p+1/2)+1\ov\sqrt{2\ta/\pi\;}})
\!-\!{\rm erf}({(p+1/2)-1\ov\sqrt{2\ta/\pi\;}}))
\end{array}
\ov
\sum\limits_{q\in Z}e^{-\pi q^2/2\ta}
}
\nonumber
\\
&{}&\!
\simeq
4*\exp(-{\pi\ov4\ta})
\label{neoc.9}
\eea
for $\ta\ll1$ and that the limit $\ta\rightarrow\infty$ for arbitrary $N_{\!f}$
takes the form
\beq
{\<\psd P_\pm \ps\>\ov(|e|/\sqrt{\pi})}=
{e^{\pm\th/\rch(\sqrt{N_{\!f}}\la/2)}\ov 4\la}
\exp\Big\{
{\ga\ov N_{\!f}}+{1\ov N_{\!f}}\log({\sqrt{N_{\!f}}\la\ov\pi})
-{2\ov N_{\!f}}\sum\limits_{j\geq1}(-1)^jK_0(j\sqrt{N_{\!f}}\la)
\Big\}
\qquad .
\label{neoc.10}
\eeq


\clearpage
\end{document}